\documentclass[12pt,preprint]{aastex}

\def\lesssim{\mathrel{\hbox{\rlap{\hbox{\lower4pt\hbox{$\sim$}}}\hbox{$<$}}}}
\def\gtrsim{\mathrel{\hbox{\rlap{\hbox{\lower4pt\hbox{$\sim$}}}\hbox{$>$}}}}

\newcommand{\beq}[1]{\begin{equation} #1 \end{equation}}
\newcommand{\beqa}[1]{\begin{eqnarray} #1 \end{eqnarray}}
\newcommand{\deriv}[2]{\frac{ d #1 }{ d #2 }}

\newcommand{\msun}{M_\odot}
\newcommand{\risco}{r_\mathrm{ISCO}}
\newcommand{\nuisco}{\nu_\mathrm{ISCO}}
\newcommand{\rhor}{r_\mathrm{hor}}
\newcommand{\hz}{\mathrm{Hz}}

\begin{document}

\title{GRMHD PREDICTION OF CORONAL VARIABILITY IN ACCRETING BLACK HOLES}

\author{Scott C. Noble, Julian H. Krolik}
\affil{Physics and Astronomy Department\\
Johns Hopkins University\\ 
Baltimore, MD 21218}

\email{scn@jhu.edu; jhk@jhu.edu}

\begin{abstract}

On the basis of data from an energy-conserving 3D general relativistic MHD
simulation, we predict the statistical character of variability in the coronal
luminosity from accreting black holes.  When the inner boundary of the corona
is defined to be the electron scattering photosphere, its location depends
only on the mass accretion rate in Eddington units $\dot m$.  Nearly independent
of viewing angle and $\dot m$, the power spectrum over the range of frequencies
from approximately the orbital frequency at the ISCO to $\sim 100$ times lower
is well approximated by a power-law with index -2, crudely consistent with the
observed power spectra of hard X-ray fluctuations in AGN and the hard states of
Galactic binary black holes.  The underlying physical
driver for variability in the light curve is variations in the accretion rate
caused by the chaotic character of MHD turbulence, but the power spectrum of
the coronal light output is significantly steeper.  Part of this contrast
is due to the fact that the mass accretion rate can be significantly modulated by
radial epicyclic motions that do not result in dissipation, and therefore
do not drive luminosity fluctuations.  The other part of this contrast is
due to the inward decrease of the characteristic inflow time, which leads
to decreasing radial coherence length with increasing fluctuation frequency.

\end{abstract}


\section{Introduction}

In a rough manner of speaking, the light from all accreting black holes, whether
those of stellar mass (Galactic Black Holes, or GBHs) or those residing in galactic
nuclei with masses $10^6$--$10^9$ times larger (Active Galactic Nuclei, or AGN),
can be divided into a thermal and a coronal portion.  The former corresponds to
the part of the continuum spectrum with a clear characteristic energy (typically
$\sim 1$~keV in GBHs, $\sim 10$~eV in AGN) and is thought to be the result of
nearly-LTE emission from the surface of the accretion disk feeding the central
black hole.   The latter corresponds to the part of the continuum that extends
approximately as a power-law from energies of order the thermal peak all the
way up to $\sim 100$~keV and is thought to arise from inverse Compton scattering
of seed photons provided by the thermal component.  Both components vary.
Compared at the same (temporal) frequency, the coronal part generally varies with
greater amplitude \citep{RMcC06} at frequencies within a few orders of magnitude
of inner disk dynamical frequencies \cite{Edelson00,Breedt09}.  It is the object
of this paper to link, for the
first time, the character of this coronal variability to heating processes directly
driven by accretion dynamics.

Coronal variability from black holes has been the subject of empirical study for
several decades.  The simplest way to characterize this phenomenon is in terms
of its Fourier power density spectrum (PDS), $P(\nu)$.  Over several orders of
magnitude in frequency, most of the observed fluctuation power is contained in
a continuum that varies smoothly with frequency.  It is convenient to describe
this continuum in terms of its logarithmic derivative with frequency $\alpha$.
With the sign convention that $P(\nu) \propto \nu^{\alpha}$, these slopes
are commonly in the range $-2 < \alpha < 0$.

GBHs move among a repertory of spectral states in which the balance
between thermal and coronal luminosity changes.  The detailed shape of $P(\nu)$
for the coronal luminosity appears to be correlated with the specific
spectral state of a GBH (as summarized in \cite{RMcC06}).
In the ``low/hard" state, $\alpha$ tends to decrease slowly with increasing
frequency: for example, in GROJ1655-40, $\alpha \simeq 0$ for
$\nu < 1$~Hz, and then gradually falls to less than $-1$ at higher frequencies,
while for Cyg X-1, it decreases from $\simeq -1.7$ to $\simeq -2.4$ over
approximately the same frequency range \citep{Rev00}.
By contrast, in the ``steep power-law" state of this system, $\alpha \simeq -1$
over almost the entire observed frequency range, from $0.01$ to $100$~Hz.  In the
``thermal" state, in which the coronal luminosity is so weak that it dominates
the spectrum only at comparatively high energies, $\alpha \simeq -1$ at
low frequencies like the steep power-law state, but steepens to $\simeq -2$
at high frequencies, somewhat resembling the low/hard state Cyg X-1, but with
much smaller amplitude variations.  Despite the coronal component's comparative
weakness in the thermal state, its variations
dominate those of the thermal component, at least for photons of more than a few
keV \citep{Chur01}.  GBHs can also exhibit quasi-periodic oscillations (QPOs):
in the low/hard state at $\sim 1$~Hz, and sometimes in the transition to the steep
power-law state at frequencies of several hundred Hz.

In AGN, whose spectral state does not appear to change in the same manner,
the power spectra may be more simply described: typically 
$-2 \lesssim \alpha \lesssim -1$ below some characteristic frequency, but
falling to $\lesssim -2$ at higher frequencies
\citep{Mark03,Mark07,Are08}.  QPOs, which, particularly in their lower
frequency variety, are easy to see in the low/hard state power spectra of GBHs,
are either entirely absent or at least difficult to detect \citep{VaUg05}; there
is at present only one reasonable candidate in an AGN \citep{Gier08}.
The variability of AGN and GBHs can be linked (approximately) through a
simple scaling: the
frequency of the roll-over in the power spectrum appears to be inversely proportional
to the mass, as the most naive theory might predict, albeit with a correction
for the accretion rate \citep{McH06}.

In order to understand the nearly featureless character of X-ray power spectra, 
many disparate theories have been proposed.  The physical content of these theories
has gradually increased over the years.  The very first ideas were purely formal:
the aperiodic nature of the 
light curves of Cyg X-1 implied to some that they were due to uncorrelated
shot noise (\cite{Terr72}).  Phenomenological models followed, as others supposed
that power spectra that were power-laws in frequency resulted from a
confluence of power-laws---in radius---that might relate the local emissivity or
propensity to hot spots to the orbital frequency \citep{K91,A91,Pech08}.
These models are consistent with the observation that many AGN are phase
incoherent \citep{1993ApJ...402..432K} and do not exhibit limit cycle behavior 
\citep{1997MNRAS.285..365C}.  The fact that there can be substantial power at
frequencies far below the orbital frequency of the inner disk coupled with
the assumption that X-rays are predominantly emitted at small radii suggested to
some that dynamics at large radii, where the dynamical timescales are longer,
control the low frequency behavior by modulating the mass accretion rate.
For example, \cite{1997MNRAS.292..679L} sought to explain the shape of the
power spectrum in terms of fluctuations at large radius in the disk's ratio of
stress to pressure, the Shakura-Sunyaev $\alpha$ parameter.  Elaborating this idea, 
\cite{Chur01} suggested that such fluctuations might explain the
connection between PDS form and spectral state in GBHs.  Alternatively,
\cite{Ax06} proposed that the changes in power spectrum might be caused by
disk precession.

None of this early work had any direct connection to physical mechanisms.  Most
importantly, in the past fifteen years we have come to realize that accretion is
driven by MHD turbulence, and the turbulence is stirred by the magneto-rotational
instability \citep{BH98}.  To move from phenomenology to physics,
models must make contact with the underlying physics of accretion.  For example,
studies of the long-term behavior of MHD turbulence might determine whether
the fluctuations in $\alpha$ suggested by \cite{1997MNRAS.292..679L} and
\cite{Chur01} actually occur.  It would also be highly desirable to
link more directly fluctuations in the accretion rate with fluctuations in
light output.  Although in the long-run the energy available for radiation is
governed by the accretion rate, time-dependent disks may have an accretion rate
that is not exactly the same at all radii, the heating rate may not exactly follow
even the local accretion rate, time is required to generate photons once gas is
heated, and the photons, once emitted, can take a finite time to make their way
out of the disk.  Indeed,
\cite{Reig06} argue on the basis of the distinct variability properties of
the coronal and thermal components in the soft state, and the correlation
between luminosity and coronal power-law slope, that fluctuations in the accretion
rate cannot on their own explain the observed variability in GBHs.

More recent work has begun to follow this path, but typically has used proxies
for the radiation rate rather than a measure of the time-varying luminosity itself.
\cite{HK01} took the first step.
They computed the power spectra of both the mass accretion rate in the plunging
region and the volume-integrated magnetic stress, with the thought that one or
the other would be a reasonable predictor of the time-dependence of the light
output.  The two power spectra were similar, but not identical, both crudely
describable as power-laws with $\alpha \simeq -1.5$.  \cite{AR03} elaborated
this approach.  Assuming that the {\it local} emissivity follows directly
from the {\it local} mass
accretion rate (and not separating the coronal part from the thermal part),
they used the vertically-integrated and azimuthally-averaged magnetic
shear stress from 3D pseudo-Newtonian MHD simulations as a proxy for the local
accretion rate.  Placing the resulting emissivity in the disk's equatorial plane
and assuming further that the fluid followed circular orbits,
they calculated the light curves seen by distant observers, allowing for general
relativistic ray-paths and Doppler-shifting.
Even though the PDSs of individual radial annuli were well described by broken
power-laws ($\alpha_1 = -1$ and $\alpha_2 = -3.5$) whose breaks were near the local
orbital frequency, the superposition of these PDSs---because of the radial dependence
of the emissivity---led to $\alpha=-2$ power-law PDSs in the total output.
\cite{MaMa04} came to similar conclusions based on a Fourier analysis
of the mass accretion rate in the plunging region.
Moving slightly closer to incorporating radiation mechanism physics,
\cite{2006ApJ...651.1031S} used data from 3D MHD simulations in full general
relativity to predict model light curves and power spectra from the {\it thermal}
component alone.  Most recently, \cite{ReynMill09} studied the fluctuations in
a variety of dynamical quantities monitored in a 3-d pseudo-Newtonian MHD
simulation, hoping to find an origin for QPO behavior.

In this paper, we seek to connect dynamical calculations still more tightly to
radiation.  The tool we bring to bear on this problem is a new fully general
relativistic 3D MHD simulation code (described in \cite{Noble09}).  Because this
code intrinsically conserves energy, it can self-consistently relate dynamics to
heating.  However, because inclusion of simultaneous radiation transfer is not
yet feasible, we cannot provide a complete account of the radiation output.
In particular, photon diffusion times within the disk body are so long (shearing box
calculations that do include radiation transfer have shown that they are generally
$\sim 10$ orbital periods: \cite{HKS06}) that diffusion delays can substantially
affect the time-dependence of the emerging light.  Consequently, in this paper
we focus on the variability of the luminosity from the coronal region, where
optical depths are likely no more than order unity (see, e.g., \cite{Ibrag05}).

The specific example we treat is one in which the black hole rotates with spin
parameter $a/M=0.9$.  Radiation is created with an emissivity in the fluid frame
which depends on the local temperature in a way designed to give the disk a desired
aspect ratio. Any dissipated heat is radiated on an orbital timescale; in
this fashion, we attempt to make the time-dependence of the light output follow
closely the time-dependence of heat-generation by such mechanisms as shocks or
magnetic reconnection.  The time and energy at which photons arrive at infinity are
computed on the basis of fully general relativistic ray-tracing including an
allowance for all travel time effects.
 
The remainder of this paper is organized as follows:
In Section~\ref{sec:methodology} we remind the reader of the salient characteristics
of our simulation and detail the new features of our ray-tracing method.  In that
section we also define our time-series analysis methodology.
Section~\ref{sec:results} presents our results, which we discuss
in Section~\ref{sec:discussion}. 

\section{Methodology}
\label{sec:methodology}

For more than a decade, MHD turbulence driven by the magnetorotational instability
has been recognized as the prime driver of accretion \citep{BH98}.  Numerical
simulations are the most powerful tool we have for studying turbulence, and in
recent years methods have been developed that permit simulations of accretion
disks over significant radial ranges in full 3D using general relativistic
dynamics \citep{dVH03,Shafee08,Noble09}.  From these and analogous 2D simulations
\citep{GSM04,Fragile09}, a consistent picture has emerged, despite a wide
range of numerical algorithms and gridding schemes:
Most of the accreted material flows through a dense disk that 
orbits the black hole at very nearly the angular frequency of circular orbits 
in the equatorial plane.  Within this dense disk, relatively small velocity
fluctuations are superposed on the bulk's orbital motion.
Higher in latitude, the disk becomes less dense,
more magnetized, and more organized in both magnetic field and velocity.

The simulation code we used to create the data discussed here was described in
\cite{Noble09}.   It is an intrinsically conservative ideal GRMHD code called
\texttt{HARM3D} that accurately captures any gridscale numerical dissipation as
heat. Numerical dissipation in many ways emulates natural dissipation; when
shocks collide and magnetic fields reconnect, entropy is created and the gas
is heated.  Left unchecked, the continual dissipation would make much of the
disk unbound and lead to a progressively growing disk thickness.  Both to permit
creation of a (statistical) steady-state and to track the rate at which energy
is dissipated, we inserted into the stress-energy conservation equation
an artificial cooling function; i.e., this equation was given the form
$\nabla_\mu T^\mu_\nu = -{\cal L}u_\nu$, where $\nabla$ denotes a covariant
derivative, $T^\mu_\nu$ is the complete stress-energy tensor, ${\cal L}$ is
the radiative emissivity in the fluid frame, and $u_\mu$ is the fluid four-velocity.
The cooling function ${\cal L} = \Omega_K \epsilon f(T/T_*)$, where $\Omega_K$
is the local Keplerian frequency, $\epsilon$ is the proper thermal energy density,
and $f(T/T_*)$ is a continuous function that is zero for $T/T_* < 1$ and increases
at higher temperatures.  The local target temperature $T_*$ is a function of radius
chosen to regulate the disk to a nearly constant aspect ratio $H/r$; in the
simulation discussed here, $H(r)/r \sim 0.05-0.12$.  Only gravitationally bound
material is cooled, and (as suggested by the form of our stress-energy equation),
the radiation is assumed to be isotropic in the fluid frame.  This simple 
radiation model was used because we are primarily interested in the bolometric 
emission from the disk and wish to apply it to a wide variety of black hole systems.
A more model-dependent cooling function could also be used \citep{Fragile09}, but it
would be computationally more expensive and would also require choosing both
a specific black hole mass and an accretion rate.

Our numerical domain was divided into $192\times192\times64$ cells in the radial, 
poloidal, and azimuthal directions respectively, with 
$r \in [1.28,120]M$, $\theta \in [0.05\pi,0.95\pi]$,
$\phi \in [0,\pi/2]$\footnote{Note that throughout this paper we use geometrized
units with $G=c=1$ unless mentioned otherwise; distances and times are therefore
scaled to the mass of the black hole $M$.}.  The radial discretization is
logarithmic---$\Delta r \propto r$---to resolve
finer features at smaller radius.  The azimuthal resolution is constant, 
and the poloidal discretization is rarefied at the poles and concentrated at the 
equator.  

The pressure maximum of the initial distribution---at $r=25M$---sets the 
location  within which a well defined accretion flow exists.  The disk reaches
an inflow steady-state for $r \leq 14M$ over the period $t=[7000M,15000M]$;
we examine only this epoch here.  For reference, the orbital period at radius $r$ is 
$T_\mathrm{orb}(r) = 3.1\times10^{-4} \left(M/10\msun\right)\left[ \left(r/M\right)^{3/2} + a/M \right]$~s.
The span $\Delta t = 8000M$ represents approximately 
$287$ orbital periods at the innermost stable circular orbit (ISCO) and $10$ orbital
periods at the initial pressure maximum.  For our black hole 
spin parameter, $\risco=2.32M$ and the horizon is located at 
$\rhor = 1.44 M$.  The disk's rest-mass density $\rho$, $4$-velocity $u^\mu$ and 
cooling function $\mathcal{L}$ evaluated at all grid points are written to disk 
every $20M$ in time.  We use this data as input to our radiation
transfer procedure to create light curves.  
Any emission outside $r=25M$ is ignored.

Because the focus of this paper is time-variability properties, we point out that
our method has two limitations that affect the shortest timescales.  Sampling at
intervals of $20M$ means that no frequencies higher than $1/(40M)$ can be probed;
our Nyquist frequency is 0.7 times the orbital frequency at the ISCO.  The other
limitation comes from our cooling rate.  Because the characteristic cooling rate
is $\sim \Omega_K$, heating fluctuations on timescales shorter than
$\sim \Omega_K^{-1}$ cannot be translated into equally rapid emission fluctuations,
even though some cooling mechanisms, notably inverse Compton scattering, can often
have cooling rates considerably faster than $\sim \Omega_K$.
In sum, we cannot present results on frequencies above $\simeq 0.7 \nu_{ISCO}$,
and the form of our cooling function potentially suppresses some fluctuation
power at the higher frequency end of the range we do discuss.

\subsection{Radiation Transfer} 

Within the simulation, we do not consider any interaction between the emitted
radiation and the material.  However, more realistically, there is always some
opacity, and in most circumstances the dominant opacity in the material near
a black hole is electron scattering.  
This opacity leads to a natural division of the radiation in two parts: that
emitted inside or outside the photosphere.  Within the
photosphere, scattering can add substantially to the time a photon can take
to reach the outside, washing out fluctuations in intrinsic emissivity; outside
the photosphere, of course, scattering has very little effect on photon escape
time.  In addition, photons deriving their energy from dissipation inside and
outside the photosphere can be distinguished spectrally: Inside the photosphere,
thermalization is strong, and the local spectrum should be approximately black
body, at a temperature $\sim 10$~eV in AGN, $\sim 1$~keV in GBHs.  By contrast,
outside the photosphere, much lower gas densities and much higher ratios of
heating density to mass density lead to much higher temperatures, and the
primary emission mechanism is inverse Compton scattering,
so that the radiated spectrum is characteristically a power-law extending well into
the hard X-ray regime.  In order to make a realistic estimate of the light curve
directly from the simulation's emissivity data, we therefore restrict our efforts
to the coronal hard X-ray emission, whose source is near or outside the scattering
photosphere.

To locate that photosphere requires a calculation of the opacity, yet
its magnitude is not defined in code-units because the simulation requires
no absolute density scale.  Instead, we determine
it after the fact by the following procedure:  We distinguish quantities in
code-units from quantities in physical units by attaching a subscript $c$ to
the former, and leaving the latter unlabeled.  If a fraction $\eta$ of 
the rest-mass of accretion were transformed into luminosity at infinity, it would be
\begin{equation}
L = \eta \int \, d\theta d\phi \sqrt{-g} \rho_c u^r (\rho/\rho_c)
(GM/c^2)^2 c^3
\ = \  \eta \dot{M}_c \, (\rho/\rho_c) (GM/c^2)^2 c^3
\label{luminosity-scale}
\end{equation}
because the unit of length is $GM/c^2$ if $G=c=1$, and $u^\mu$, when measured in
units of $c$, is dimensionless.  Here, $g$ is the determinant of the metric. 
Normalizing the luminosity to the Eddington
luminosity $L_E$, we find that the relation between physical density and code
density is
\begin{equation}
\rho/\rho_c = \frac{4\pi c^2}{\kappa_T G M \dot{M}_c} \, \frac{L}{\eta L_E },
\label{density-scale}
\end{equation}
where $\kappa_T$ is the electron scattering opacity per unit mass, and
$\dot{M}_c=0.0177$ is the time-averaged rest-mass accretion rate in code units.  
By fortunate coincidence, optical depths depend {\it only} on $L/(\eta L_E)$,
which we abbreviate as $\dot m$, because the unit of length is $\propto M$.

Because our accretion flow is far from spherically symmetric, the location of the
photosphere is a function of the observer's position.  We imagine, then, that
numerous ``cameras" are placed on a grid in polar angle $\vartheta$ and azimuthal
angle $\varphi$ on a very large sphere (radius $10^6M$) centered on the black hole.
From each camera, we define a bundle of geodesics that run through the problem
volume.  These are parameterized by an affine parameter $\lambda$ normalized so
that an observer in the local fluid frame would measure the differential length
along a ray as
\begin{equation}
ds = \nu d\lambda,
\end{equation}
where $\nu$ is the frequency of the photon as measured by that observer.
If $N^\mu = dx^\mu/d\lambda$ is the $4$-vector tangent to the null ray then 
\beq{
\nu = \frac{\nu_\mathrm{cam}}{z} \label{redshift}
}
where $z$ is the redshift factor between the local fluid frame observer and
the camera frame:
\begin{equation}
z = \frac{\left(u^\mu N_\mu \right)_\mathrm{cam}}
               {\left(u^\mu N_\mu \right)_\mathrm{ff}}\quad . \label{redshift-factor}
\end{equation}
In the numerator of this ratio, the 4-velocity is that of the camera; in
the denominator, it is that of the fluid at some point along the ray.
We then integrate the optical depth
\begin{equation}
d\tau  = \rho \kappa_T \nu d\lambda \label{optical-depth}  
\end{equation}
along these geodesics in order to determine the location of the photospheric
surface for that camera.  The photosphere surface is defined to lie
at a constant $\tau = \tau_\circ$, which we set to unity, i.e., $\tau_\circ=1$. 

Once the location of the photosphere is determined, we integrate the emissivity
along these geodesics from the photosphere out to the camera; we assume no
scattering takes place along these rays:
\beq{
\deriv{}{\lambda}\left(\frac{I_\nu}{\nu^3}\right) = 
\frac{j_\nu}{\nu^2}  \quad ,  \label{gr-rt-eq}
}
where $I_\nu$ is the specific intensity and $j_\nu$ is the fluid-frame
emissivity, given by:
\beq{
j_\nu = \frac{\mathcal{L}}{4 \pi} \, \delta\left(z \nu - \nu_\mathrm{cam}\right) 
\quad . \label{emissivity}
}
Integrating over all $\nu_\mathrm{cam}$ to find the bolometric luminosity is
equivalent to setting $\nu_\mathrm{cam}=1$ and $\nu = 1/z$; the latter
procedure is done in practice.  To set the units of the observed luminosity, we
note that the units of power density are the units of energy density ($\rho c^2$)
divided by the unit of time ($GM/c^3$).  The end result is
\begin{equation}
{\cal L} = \frac{4\pi c^7 \dot{m}}{ \kappa_T \left(GM\right)^2 \dot{M}_c }  \quad .
\end{equation}
However, these units are also unnecessary because all our results for variability
will be shown in fractional terms, relative to the mean luminosity.

We are therefore left with three parameters to explore:
$\dot{m}$, $\vartheta$ and $\varphi$.  
We vary $\dot{m}$ from a value low enough that the entire flow is optically thin up to
the Eddington limit: 
\beq{
\dot{m} \in  \left\{ 0.001 , 0.003 , 0.01  , 0.03  , 0.1   , 0.3   , 1. \right\}
\quad . \label{mdot-values}
}
The simulation should not be biased toward any particular pole, so 
we sample $\vartheta$ only over one hemisphere, uniformly 
in $\sin\vartheta$: 
\beq{
\vartheta \in \left\{  5^\circ , 17^\circ , 29^\circ , 41^\circ , 53^\circ , 65^\circ , 77^\circ , 89^\circ  \right\}
\quad . \label{theta-cam-values}
}
Similarly, the physics of our accretion disk has no special azimuthal 
orientation, so any observed dependence of the light curves on $\varphi$ must be only 
statistical fluctuations.
However, our simulation domain spans only the first quadrant in azimuth,
from $0$ to $\pi/2$.

\begin{figure}
\epsscale{0.8}
\plotone{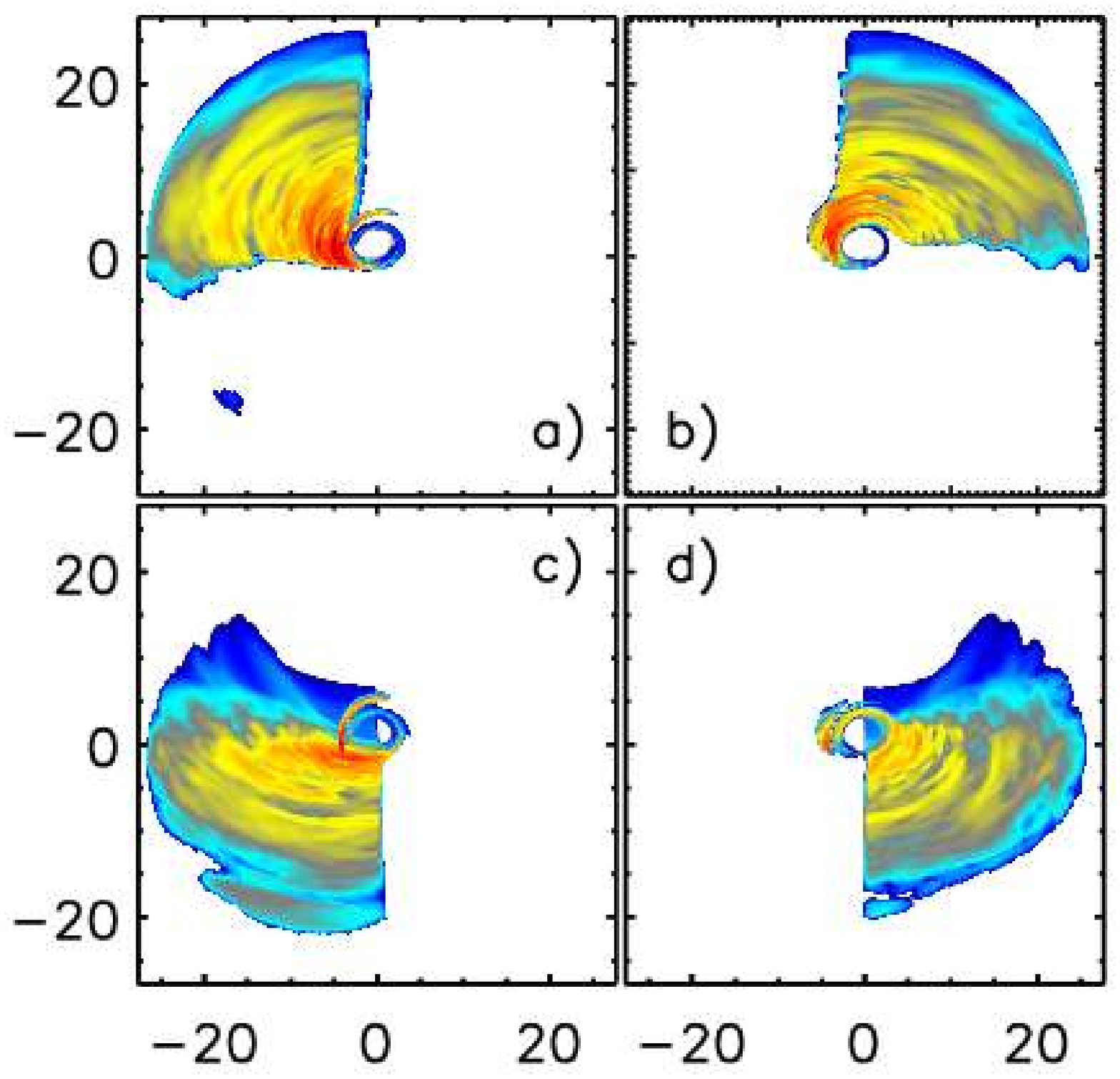}
\epsscale{0.1}
\plotone{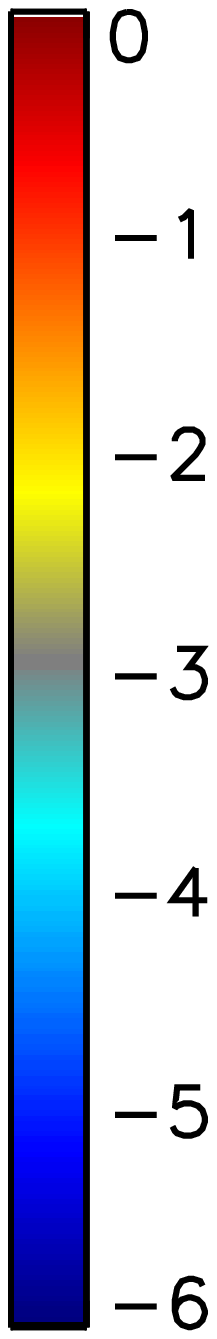}
\epsscale{1.}
\caption{(a-d) Snapshots 
at---respectively---$\varphi = \left\{\pi,\pi/2,3\pi/2,0\right\}$.  Each 
image was taken at $\vartheta=53^\circ$ and $t = 9000M$ with 
$\dot{m} = 0.01$.  The axes mark the image's vertical and horizontal 
extent in the image plane in units of $M$. (Far Right) Logarithmic color map 
used to make the images. The intensity is normalized to the maximum intensity 
of the composite image shown in Figure~\ref{fig:image-composite}. 
\label{fig:images-phi}}
\end{figure}

\begin{figure}
\epsscale{0.8}
\plotone{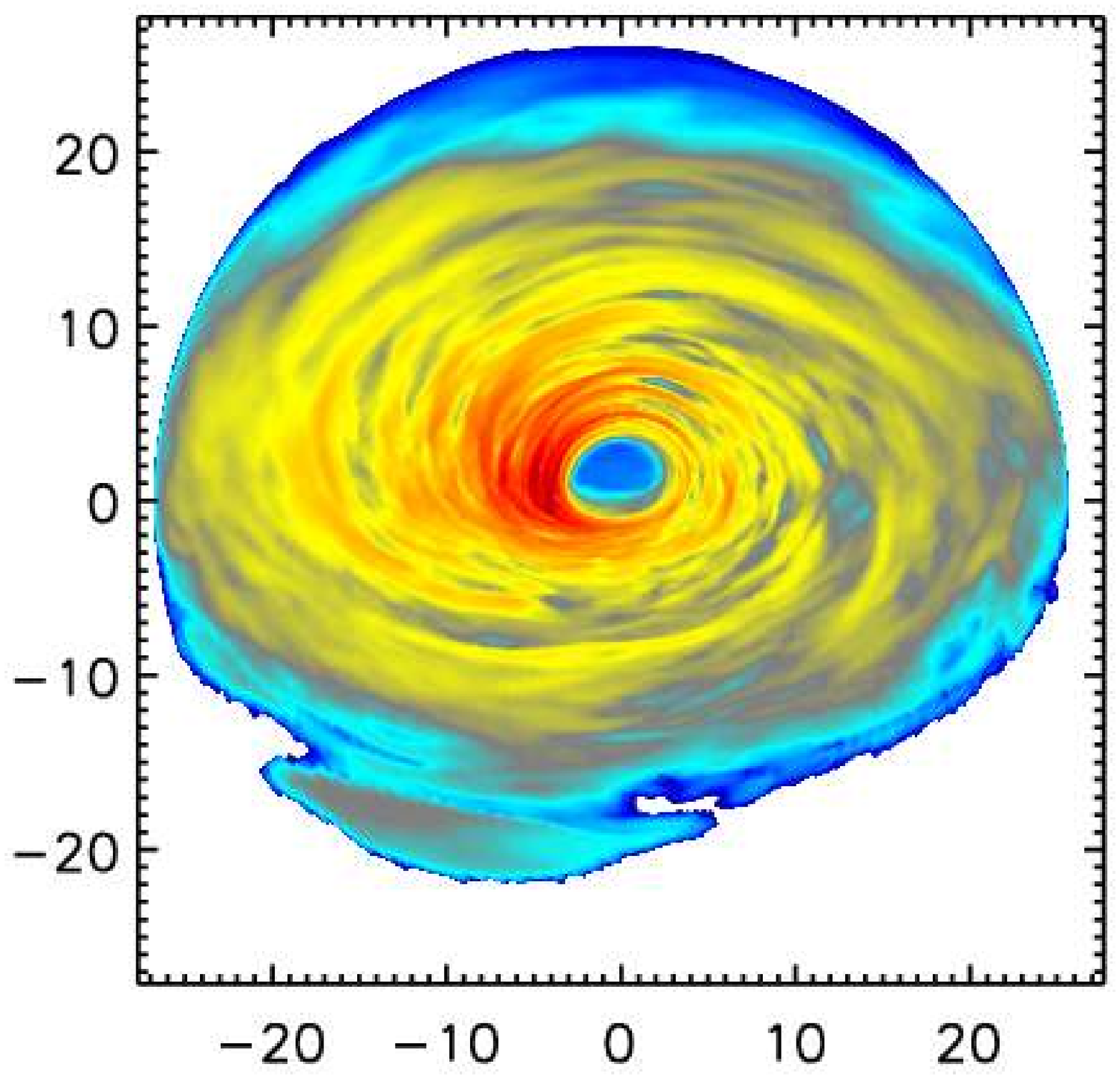}
\epsscale{0.1}
\plotone{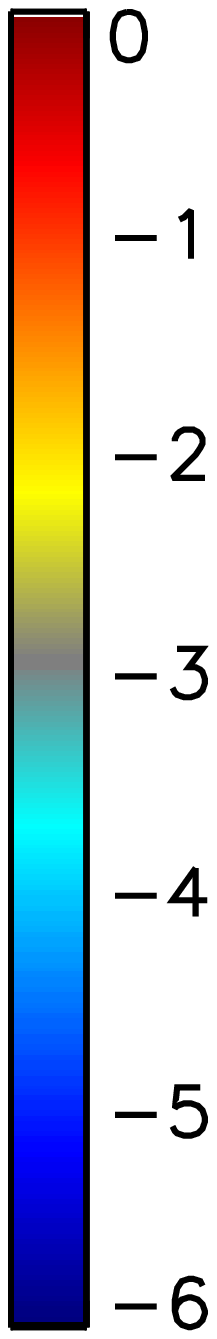}
\epsscale{1.}
\caption{(Left) Composite of the snapshots shown in Figure~\ref{fig:images-phi}. 
(Right) Logarithmic color map used for the image.  The maximum intensity
in the map has value unity.
\label{fig:image-composite}}
\end{figure}

To cope with this limitation, we remap the density and velocity data into the other
quadrants, but not the emissivity.  By doing so, we can compute the portion of
the light reaching infinity from this quadrant alone with a proper allowance for
optical depth effects in all directions.  In principle, there are four different
ways we might have placed the radiating quadrant with respect to the quadrants
having only opacity.  From the expectation of azimuthal symmetry, it then follows
that a full description of the statistical character of the light curve can be
obtained from viewing this quadrant from only four azimuthal directions,
which we choose as  
\beq{
\varphi \in \left\{  0, \frac{\pi}{2}, \pi, \frac{3\pi}{2}  \right\}
\quad . \label{phi-cam-values}
}
This piecewise construction of the composite image is illustrated in 
Figures~\ref{fig:images-phi}~-~\ref{fig:image-composite}.  Images from 
different $\varphi$ are shown in Figure~\ref{fig:images-phi}, and 
their sum is plotted in Figure~\ref{fig:image-composite}.   
We find that frame-dragging of photons can cause emission from one 
quadrant of the simulation domain to spread into all quadrants of 
the image plane.  The total flux of a quadrant varies
with $\varphi$ since certain values orient the disk's orbital 
velocity closer to the line of sight.  Figures~\ref{fig:images-phi} and
\ref{fig:image-composite}  will be helpful references in our later
discussion about how the artificial azimuthal symmetry
condition influences our variability predictions in Section~\ref{sec:azim-symm-cond}.

Tracking the light through the simulation data is complicated by the fact
that spacetime's curvature means that a set of photons reaching an observer at
one instant may originate from the accretion flow over a wide range of coordinate
times.  \cite{AR03} included the effect of time delays, but used an infinitesimally
thin emission region which intersected each  geodesic only once. 
This meant that they did not have to propagate the light rays through the 
simulation data, as we must because our emission region is extended.

There are many ways to go about the bookkeeping inherent to this problem, 
but different algorithms place vastly different demands on computer memory.
Ray-tracing as the simulation runs is expensive, 
memory intensive, and not amenable to exploration since adjustments made to the 
ray-tracing scheme would require rerunning the simulation\footnote{See 
\cite{2008arXiv0802.0848D} for an algorithm that performs the ray-tracing in situ.}. 
For this reason, we adopted a post-processing procedure, which means that our
time resolution is set by the rate at which we output simulation data.
As a ray traverses spacetime, we use quad-linear (linear in space and time)
interpolation to determine the necessary
quantities along the ray's path.  This interpolation is done with pairs of simulation
data slices at a time in order to reduce the calculation's memory footprint. 
Rays are organized into stacks of snapshots, each representing a batch of 
rays distributed over the image plane that reach the observer synchronously.   
Each snapshot then depends on a finite span of simulation data, or rather,
a given time slice of simulation data influences a sequence of snapshots. 

In \cite{Noble09}, we spent much effort to ensure our ray tracing 
calculation was converged with respect to the number of light rays used per image; 
following the analogy with a pinhole camera, we will refer to them as \emph{pixels}
in our camera.  We found best performance using a nonuniform pixelation that 
approximates a projection of the simulation's grid onto the image plane. 
Using the radial profile of flux at infinity $dF/dr$ as our 
convergence criterion, 
we found that $N_\mathrm{pix}=255^2$ nonuniform pixels and 
$N_\mathrm{pix}=1200^2$ uniform pixels result in approximately the same 
level of accuracy (largest relative error over $r$ is $5\%$) when compared
to a calculation with $N_\mathrm{pix}=1500^2$ uniform pixels. 

\subsection{Time Series Analysis} 
\label{sec:time-series-analysis}

The radiative transfer calculation yields a function 
$F\left(\vartheta,\varphi,t,\dot{m}\right)$ which represents the bolometric 
flux for an observer at constant radius ($r=10^6 GM/c^2$) and time $t$.
In the following analysis unless explicitly stated, we integrate $F$ over
$\varphi$ to yield a light curve for a full $2 \pi$ disk.  Our goal is then to
understand the temporal character (e.g., power spectrum, variance)
of this function with respect to the remaining parameters 
(i.e. $\vartheta,\dot{m}$).  

Given a function defined 
on a set of $N$ discrete times $F(t_n)$ (where $n = 0, \ldots, N-1$), 
we define its discrete Fourier transform $\hat{F}(\nu)$  at 
frequency $\nu$ as 
\beq{
\hat{F}(\nu) = \frac{1}{N} \sum^{N-1}_{n=0}  F(t_n) e^{-2 \pi i \nu t_n }
\label{discrete-fourier-transform}
}
where $\nu \in \{1/T,\ldots,N/2T\}$, and $T = t_{N-1} - t_0$.
As remarked earlier, our Nyquist limit is 
$\nu_\mathrm{max} = (40M)^{-1} \simeq 0.7 \nuisco$, where $\nuisco$
is the orbital frequency at the ISCO.  

We define the normalized PDS of the light curve $F(t)$
such that it is a measure of the fractional variance per unit frequency:
\beq{
P(\nu) = \frac{2T}{\bar{F}^2} \left| \hat{F}\left(\nu\right) \right|^2 
\quad , \label{power-spectrum-density}
}
where $\bar{F}$ is the time average of $F(t)$
\footnote{
These power spectra accurately describe our finite duration simulation.  However, as
pointed out by \cite{1982ApJ...261..337D} and \cite{1984ApJ...281..482D}, 
power spectra with slopes
steeper than $-2$ may be subject to artificial leakage of fluctuation power from low
frequencies to high.  This problem can be significant when the power spectrum is
substantially steeper than $-2$ and extends to frequencies significantly lower than those
probed by the experiment (i.e., much lower than the inverse of its duration).  Because
the slope we measure is only slightly more negative than $-2$ over most of the parameter
space of interest, we believe that this artifact may have only limited effect, but only much
longer simulations can definitively answer this question.
}.  Since $\hat{F}\left(\nu\right)$
is a complex number, we can represent it in terms of its magnitude
and phase:
$\hat{F}\left(\nu\right) = A\left(\nu\right)  e^{i \psi\left(\nu\right)}$.

Observed light curve power spectra are often described in terms
of a best-fitting power-law; as we will see, our results resemble power-laws
at about the same level of accuracy.  Determining the best-fit slope to the
power spectra of our light curves, however, is somewhat arbitrary because we do
not know the ``error distribution" of our data (the relevant ensemble for
us would be a large set of simulations begun with slightly different initial
conditions).  Because our goal is only
qualitative description, we choose a very simple approach: equal standard
error at each frequency point. That is, to find the best-fit power-law slope
for a given power spectrum, we minimize the quantity
\begin{equation}
\chi^2_{\rm eff} = \sum_i \left[P(\nu_i) - C\nu_i^{\alpha}\right]^2
\end{equation}
with respect to $C$ and $\alpha$.  This form tends to weight most heavily
the highest decade of frequencies because the frequency points are separated
by a constant $\Delta \nu = (8000M)^{-1}$.

A key physical question we wish to explore is: how well correlated is the
variability at different radii?  Answering this question will help us understand 
the light curve's power spectrum and provide a context for comparison to other
models (e.g.,\cite{1997MNRAS.292..679L}).  Correlations are performed by first
decomposing the light curve's PDS into partitions.  Let 
there be $N$ partitions, each with a different light curve $F_n\left(t\right)$ 
so that our total light curve is their simple sum:
$F\left(t\right) = \sum_{n=1}^N F_n\left(t\right)$. 
The total power spectrum $P(\nu)$ can therefore be expressed as a sum of the
partitions' PDSs ($S_a$) plus a sum that depends on how well the different
modes are correlated ($S_b$):
\beqa{
P(\nu) & = & \frac{2T}{\bar{F}^2} \left| \hat{F}\left(\nu\right) \right|^2  
= \frac{2T}{\bar{F}^2} \left( \sum_{n} \hat{F}_n\left(\nu\right) \right) \left( \sum_{m} \hat{F}_m\left(\nu\right) \right)^* \label{power-decompostion1} \\
& = & \frac{2T}{\bar{F}^2} \left[ \sum_{n} A^2_n +  2 \sum_m \sum_{n > m} A_n A_m \cos\left(\Delta\psi_{n m}\right) \right] \label{power-decompostion2} \\
& = & \sum_{n} \frac{\bar{F}^2_n}{\bar{F}^2} P_n +  \frac{4T}{\bar{F}^2} \sum_m \sum_{n > m} A_n A_m \cos\left(\Delta\psi_{n m}\right) \label{power-decompostion3} \\
& = & S_a + S_b .\label{power-decompostion4}
}
Here $\Delta \psi_{n m} = \psi_n - \psi_m$ is the difference in phase at frequency
$\nu$ between two partitions.  Note that even though $P(\nu)$ is independent of our
partition scheme,  the relative sizes of $S_a$ and $S_b$ are not.  If the
$A_n$ are all of similar magnitude, then $S_a / S_b \rightarrow 0$ as
$N \rightarrow \infty$ ($S_b \sim N\left(N-1\right)$ while $S_a \sim N$) and 
$S_b / S_a \rightarrow 0$ as $N \rightarrow 1$ (there is only one partition and
one signal is perfectly coherent with itself).  If the partitions are perfectly
incoherent, then $P \simeq S_a$.  Conversely, 
if they are perfectly coherent, then $P \simeq S_b$ for $N>2$. 

\section{Results}
\label{sec:results}

\subsection{Light Curves and Power Spectra: Dependence on Accretion Rate
and Inclination}
\label{sec:depend-accr-rate}

\begin{figure}
\plottwo{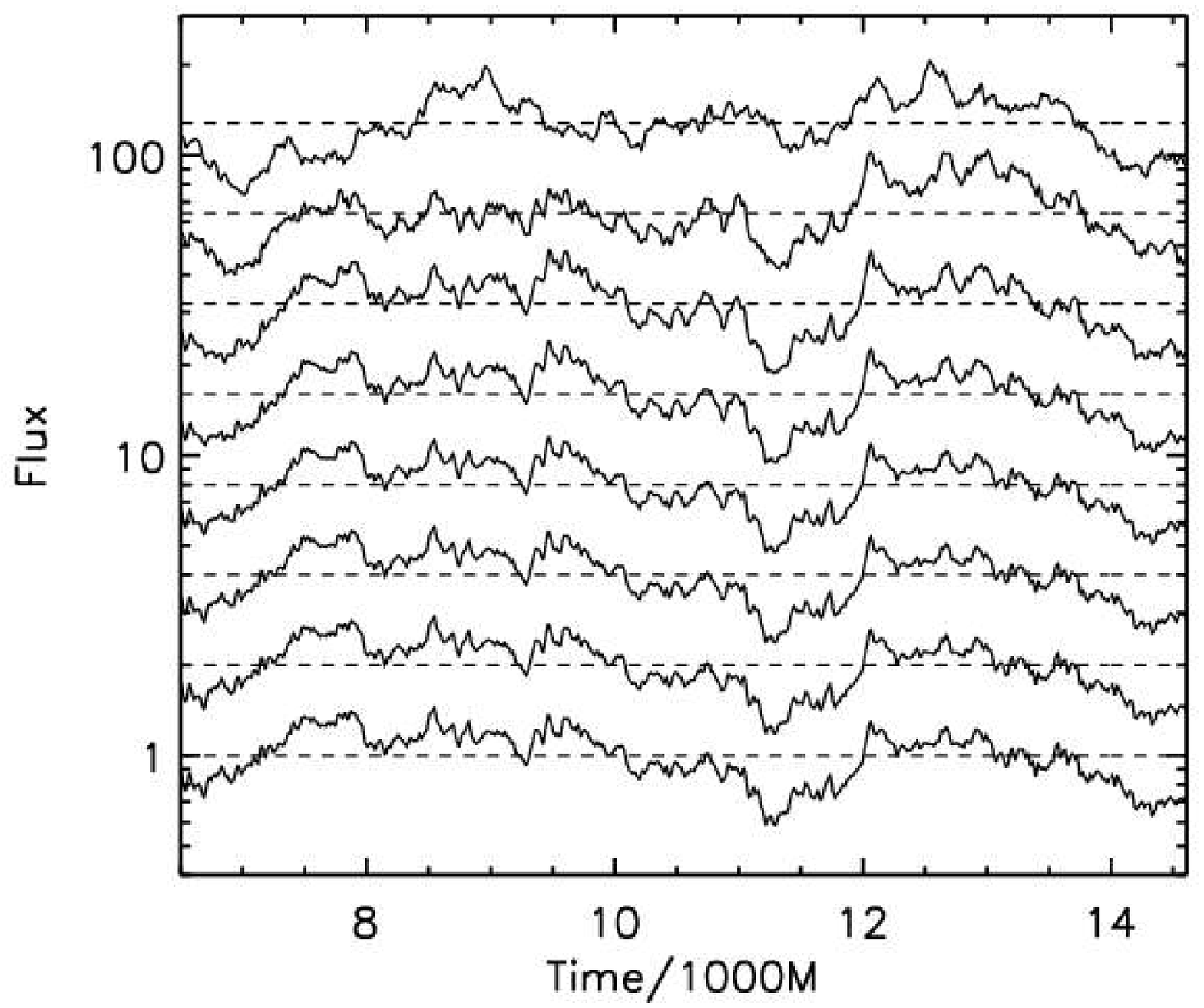}{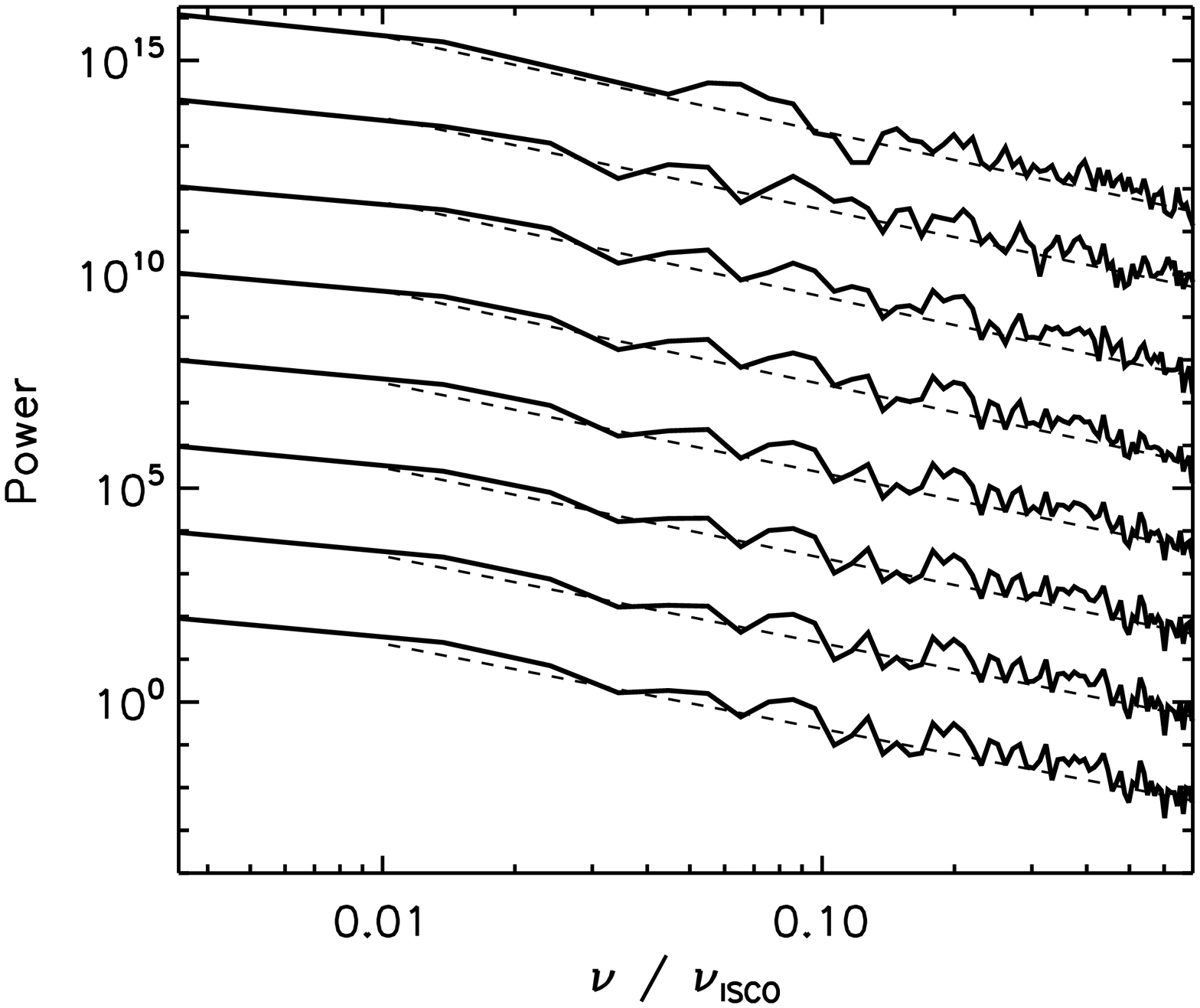}
\caption{(Left) Normalized light curves (solid curves) for $\dot{m} = 0.01$ 
and all values of $\vartheta$. The light curves and their mean values
(dashed curves) have been shifted vertically by incremental factors of two
for clarity.  (Right) Normalized power spectra of these light curves compared
to their best power-law fits (dashed curves).  The power spectra are separated
by incremental factors of $100$.  In both plots, the curves are ordered bottom-to-top
in increasing order of $\vartheta$.
\label{fig:light-curves-incl}}
\end{figure}

\begin{figure}
\plottwo{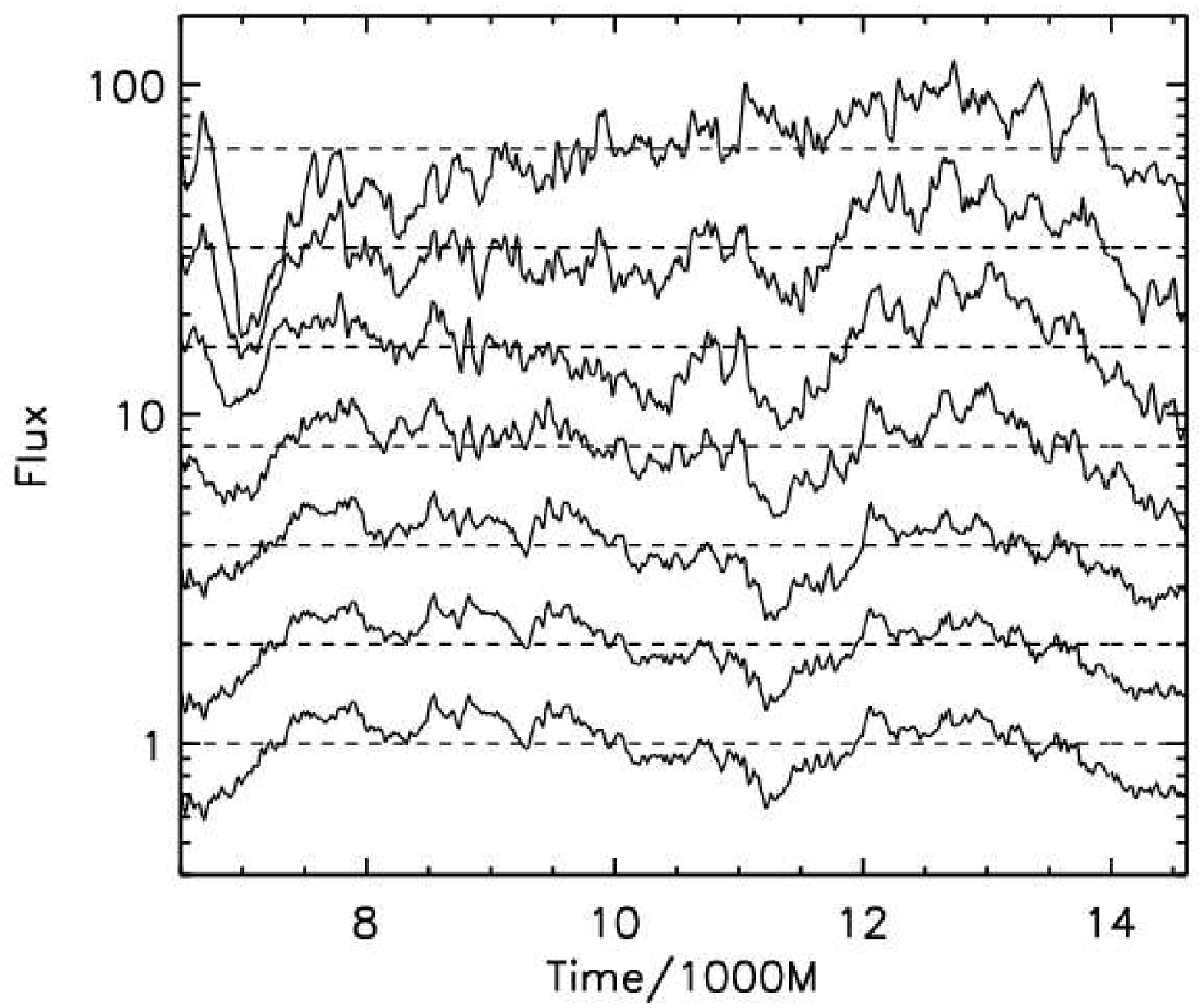}{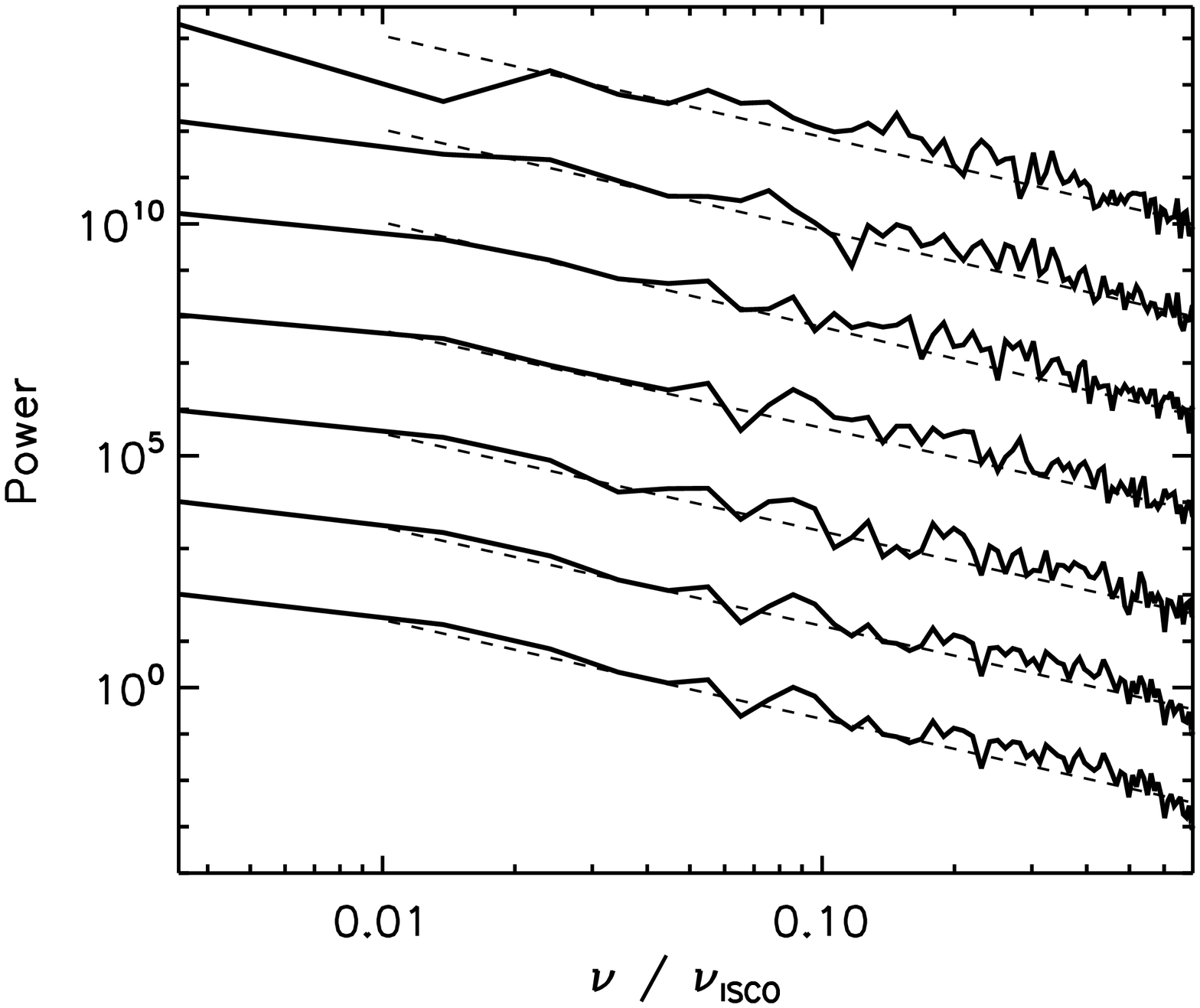}
\caption{(Left) Normalized light curves (solid curves) for $\vartheta=29^\circ$
and all values of $\dot{m}$.  The light curves and their mean values (dashed
curves) have been shifted vertically by incremental factors of two for clarity. 
(Right) Normalized power spectra of these light curves compared to their 
best power-law fits (dashed curves).  The power spectra are separated by incremental 
factors of $100$.  In both plots, the curves are ordered bottom-to-top
in increasing order of $\dot{m}$. 
\label{fig:light-curves-mdot}}
\end{figure}

\begin{figure}
\plotone{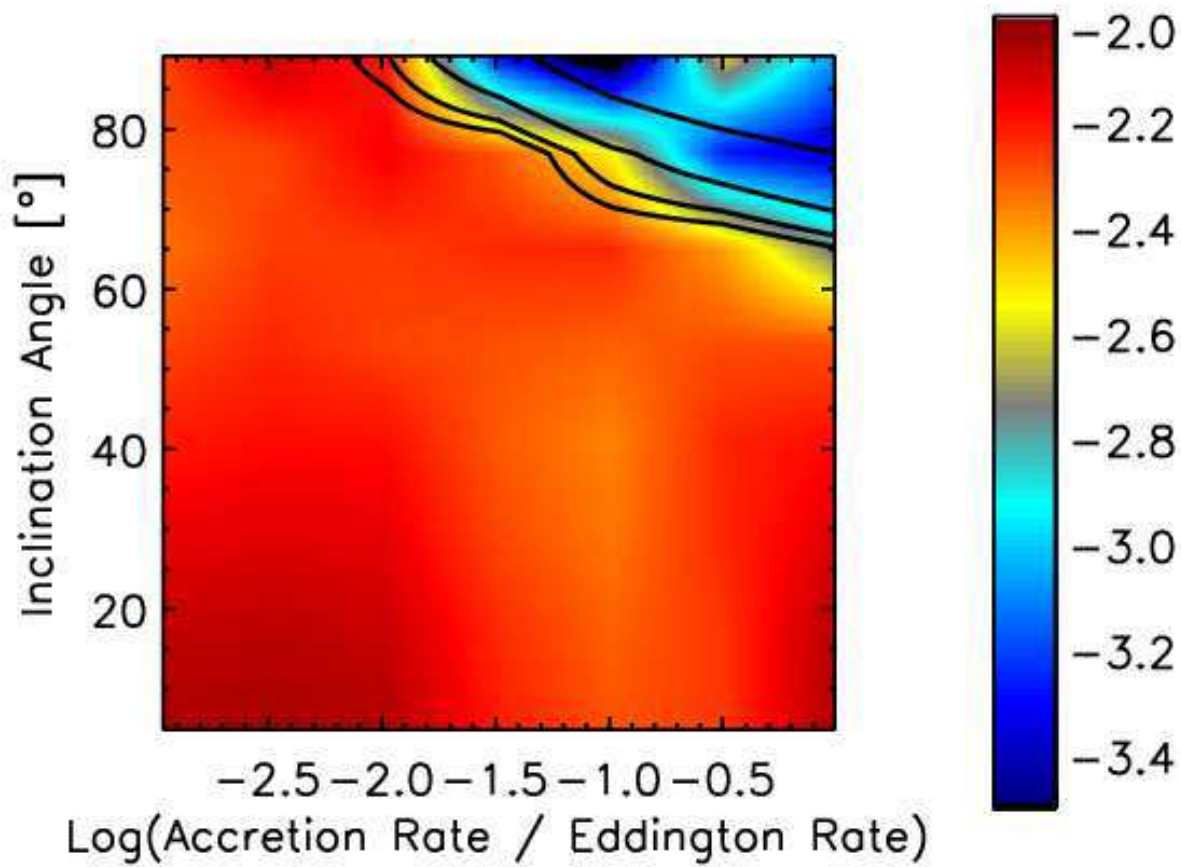}
\caption{Exponents $\alpha$ of the best power-law fits to the power spectra
of the light curves
as functions of $\dot{m}$ (horizontal axis) and inclination angle (vertical axis).
The departures from $\alpha \approx -2$ in the upper-right-hand corner of the
plot are caused by the disk's self-obscuration.  The black curves
represent contours of $R_o$.  From bottom to 
top, $R_o(\vartheta,\dot{m}) = \left\{ \risco, 3.5M, 6M, 12M\right\}$.
\label{fig:power-law-exponents-pspace}}
\end{figure}

Sample light curves and their corresponding power spectra can be seen
in Figures~\ref{fig:light-curves-incl} and \ref{fig:light-curves-mdot},
the former displaying how they change with viewing angle $\vartheta$
at fixed $\dot m$, the latter how they change with accretion rate $\dot m$ at
fixed $\vartheta$.  Changing
inclination does relatively little to alter long time-scale fluctuations, but
can lead to differences on short time-scales.  On the other hand, changing
the opacity can lead to substantial differences in the light curve even
while the viewing angle remains the same.  Remarkably, however, the gross
shape of the power spectrum is almost invariant to both sorts of changes:
the best-fit power-law index is $\simeq -2$ for all but the highest
accretion rates and inclinations (Fig.~\ref{fig:power-law-exponents-pspace}).

The strongest effects influencing the inclination dependence of variations are
relativistic beaming and boosting, which become more important as the orbital
velocity becomes larger and more nearly parallel to the outgoing geodesics.
They therefore have the greatest effect on radiation issuing
from the smallest radii when viewed at high inclination.  Because those same
inner radii have the highest dynamical frequency, one might then expect
a boost in the high-frequency portion of the power spectrum at large $\vartheta$.
In relative terms, this does {\it not} occur---as we have seen, the slope of
the power spectrum depends only weakly on inclination, except when $\dot m$
is quite large (see further discussion later in this subsection).  Nonetheless,
although the relative variance changes little with inclination, the absolute
variance, as well as the absolute luminosity, does increase when the disk 
becomes more edge-on,
as has also been seen in previous calculations \citep{AR03,2006ApJ...651.1031S}.

Because we explore only relative variability, the absolute luminosity's
proportionality to $\dot{m}$ is irrelevant to our discussion.  The accretion rate
influences the light curves in our calculations only by setting the opacity scale.
The accretion  rate is therefore degenerate with our choice of $\tau_\circ$, and
we can speak equivalently in terms of accretion rate or optical depth.

When the opacity is dominated by electron scattering, the
disk is completely transparent for accretion rates $\dot m = 0.001$ or lower.
Increasing $\dot{m}$ moves the photosphere farther from the disk's midplane,
and emission from high latitudes becomes more dominant because our disk follows
a nearly constant $H/r$ profile.  At the same time, increasing
$\dot m$ leads to a relative suppression of light from outer radii because the
disk surface density, and hence its optical depth, increases rapidly outward. 
For this reason, the largest accretion rates select out fluctuations from the 
innermost and uppermost regions of the (bound) accretion flow. 

This pruning of the coronal volume with increasing $\dot m$
is the most likely explanation for the fact that the relative variance of
the light curves monotonically increases with 
accretion rate, from $0.04$ at $\dot{m}=0.001$ to $0.09$ 
at $\dot{m}=1$.  As the region above the photosphere shrinks in radial and
vertical extent with $\dot{m}$,  it contains fewer
independently-fluctuating volumes, so that their summed emission has larger
fractional fluctuations.

Increasing $\dot m$ also leads to greater obscuration of 
high inclination observers' views of the inner disk.  It is this
effect that explains the steepening of the best-fit power-law in the
high $\dot m$ and $\vartheta$ corner of Figure~\ref{fig:power-law-exponents-pspace}.
Since we restrict the emission to $r<25M$ while the disk matter extends
out to $r=120M$, sufficiently large accretion rate and inclination angle can
lead to complete obscuration of the emission region.  The radius within which
all emission is obscured, $R_o$, is consequently an increasing function of
inclination and accretion rate.  
Curves showing how $R_o$ depends on $\dot m$
and $\vartheta$ are also shown in Figure~\ref{fig:power-law-exponents-pspace}.
Only when $\vartheta > 60^{\circ}$ is this obscuration effect significant.  Because
the location of the obscured region is sensitive to the geometry of the 
disk---which is artificially tuned to have a near constant aspect 
ratio---the steepening of the fluctuation power spectrum due to obscuration
may be artificial.

\subsection{The Optically Thin Limit: Origin of the Variability}
\label{sec:orig-emiss-vari}

\begin{figure}
\plottwo{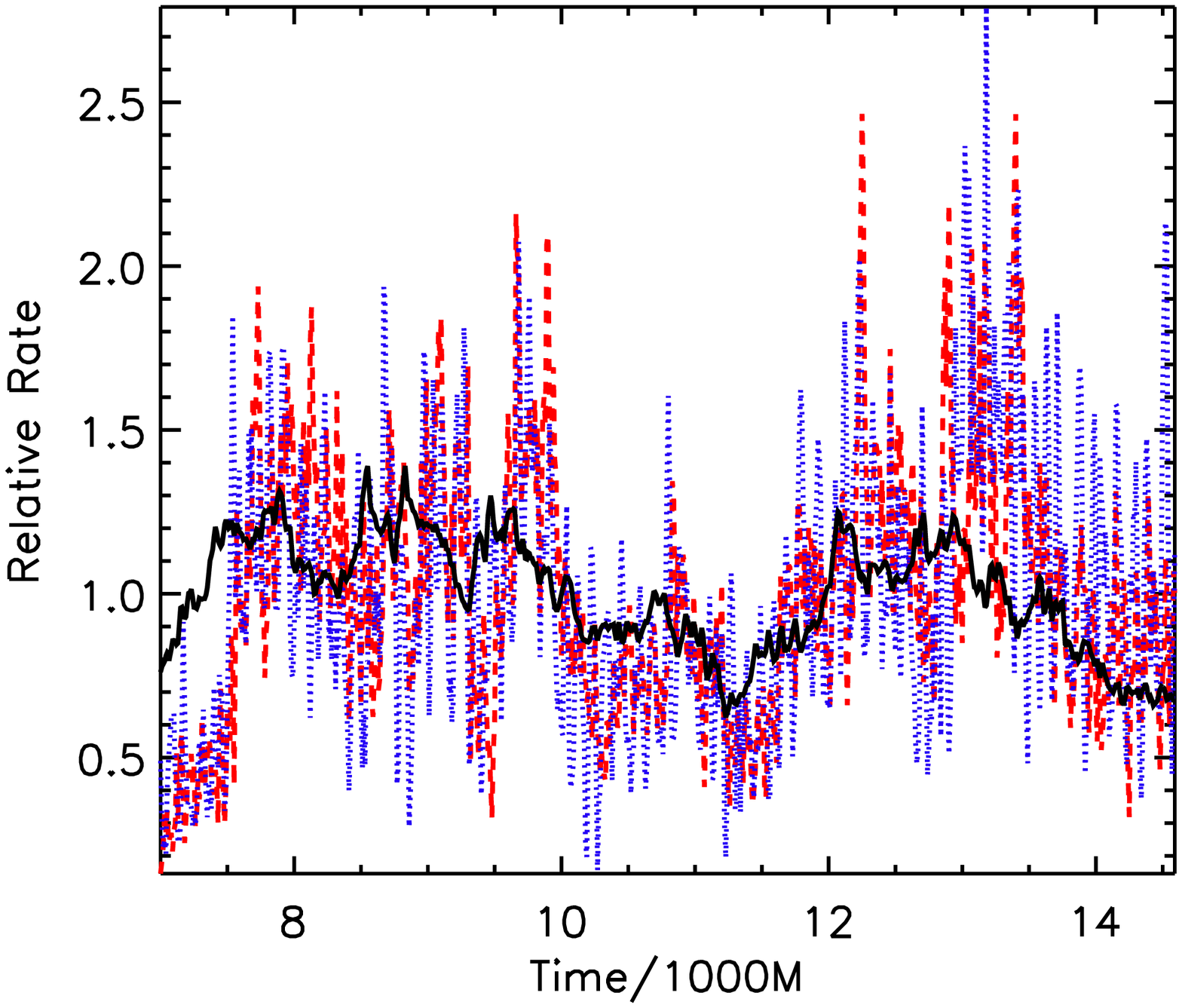}{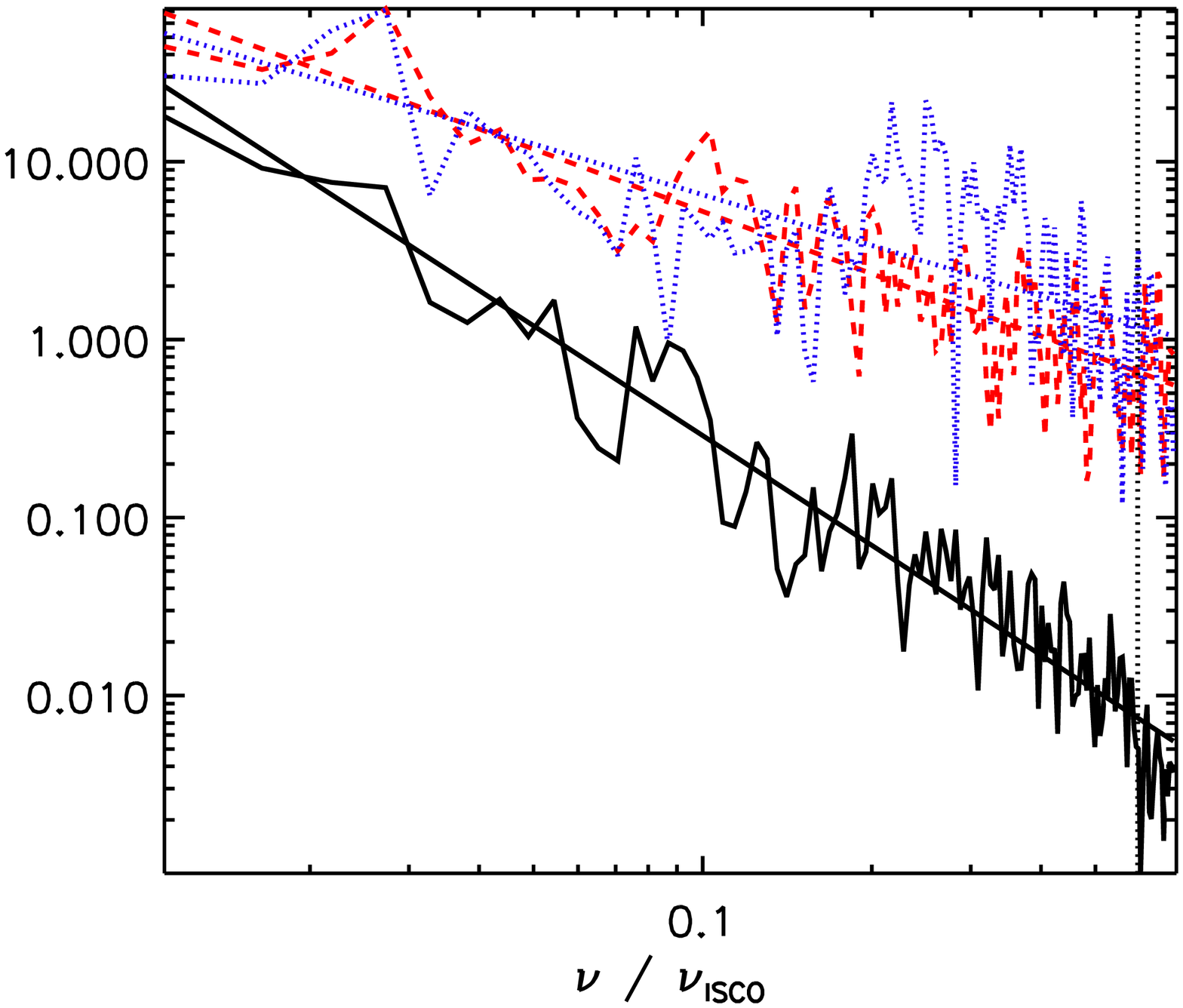}
\caption{(Left) Total flux $F(t)$ observed at $\vartheta=5^\circ$ for $\dot{m}=0.001$
(solid black curve) 
compared to the accretion rate $\dot M$ (dotted blue curve) at $r=3.5M$ and
emissivity ${\cal L}$ at $r=3.5M$ (dashed red curve).  All rates are normalized
to their time averages. (Right) Power spectra of these rates and their best-fit
power-laws.  The values of the best-fit power-law exponents are
$\alpha_{\dot{M}} = -0.9$, $\alpha_{\mathcal{L}}=-1.2$, and $\alpha_{F}=-2.1$. 
\label{fig:lightcurves}}
\end{figure}

The most obvious explanation for variability in radiative output is
variability in the 
accretion rate.  Let us first examine the relationship between the local
emissivity and accretion rate to see if this is indeed true for our simulation.
In Figure~\ref{fig:lightcurves} we compare the accretion rate and emissivity
at $r=3.5M$ to the light curve (integrated over the entire simulation)
measured face-on ($\vartheta=5^\circ$).  We choose to compare behavior at
$r=3.5M$ to the total light curve because it is the
radius of the brightest annulus, and should therefore be a major contributor
to the composite light curve.  A nearly-polar viewing angle minimizes
relativistic effects, simplifying the comparison of the observed light curve
to local emissivity.  As expected, the disk-integrated
light curve follows the same large amplitude, long timescale fluctuations seen 
in the accretion rate and emissivity at $r=3.5M$.  However, it lacks the short 
timescale variability of the local emissivity and accretion rate.  The
same effect appears, of course, in the power spectra.
At this radius (and at most of the others in the steady-state portion
of the disk), the accretion rate and emissivity power spectra are approximate
power-laws with exponents $\sim -1$ and $\sim -1.5$, respectively, significantly
shallower than the total light curve power spectrum, for which
the overall slope is $\simeq -2$.

\begin{figure}
\plotone{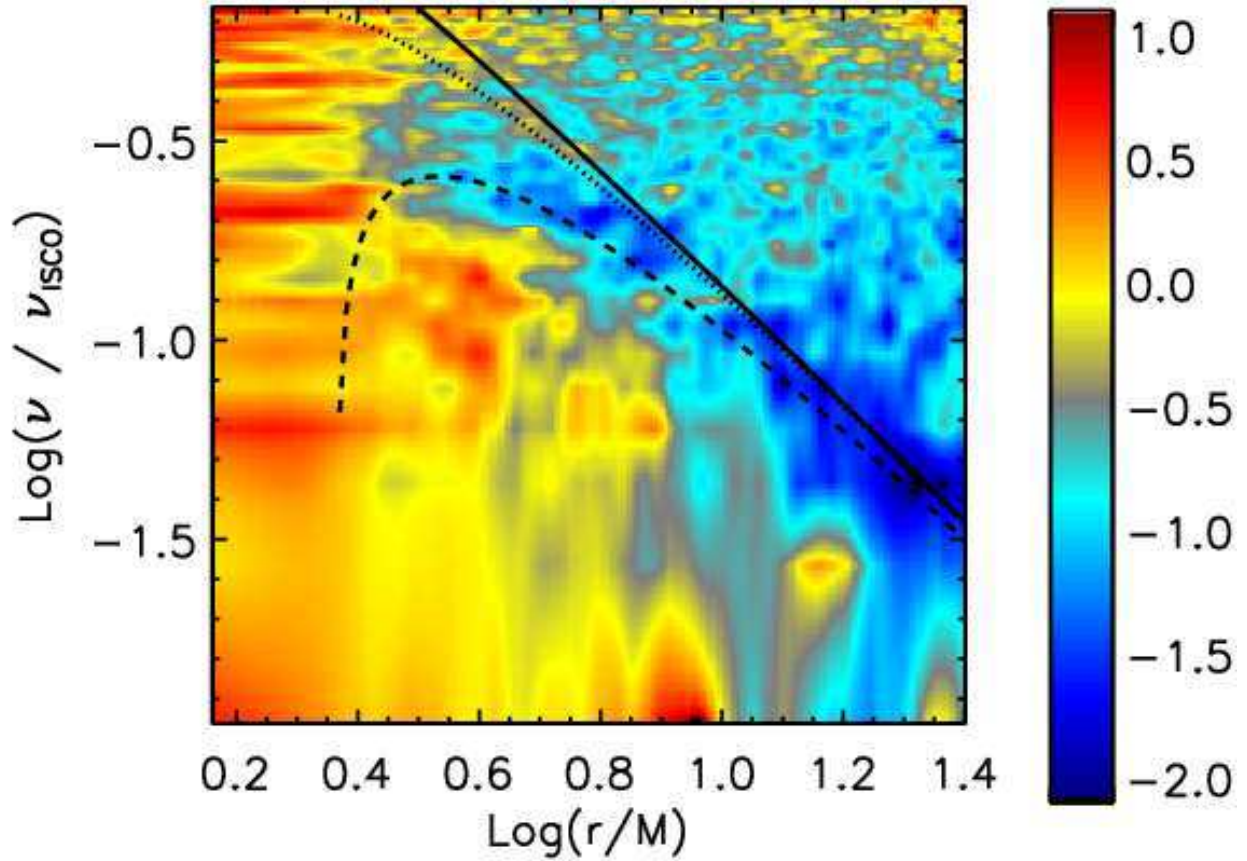}
\caption{Ratio of the emissivity's normalized power spectrum to the accretion rate's
normalized power spectrum plotted as a function of radius and frequency.  Each
power spectrum was smoothed over nine frequency bins before the ratio was taken
in order to display trends in the data more clearly.  Black curves show the orbital
frequency (solid curve), radial epicyclic frequency (dashed curve), and vertical
epicyclic frequency (dotted curve).
\label{fig:emiss-mdot-psd-ratio}}
\end{figure}

In order to elucidate why fluctuations in the local properties have more
high frequency power than the total light curve,  and to understand
better to what degree the accretion rate drives the radiation, we
examine the relationships between the power spectra of $\dot{M}(r,t)$,
${\cal L}(r,t)$, the flux from $r$ as it is observed on the polar axis
at infinity ($dF/dr$), and the total flux $F(t)$ in the polar direction.  We
have already seen that $|\hat{\dot{M}}|^2$
and $|\hat{\cal{L}}|^2$ are similar but not identical.  A closer comparison
of these two power spectra may be seen in Figure~\ref{fig:emiss-mdot-psd-ratio},
which shows the ratio of the emissivity's power spectrum to that of the accretion
rate as a function of radius and frequency.  In much of the diagram, the ratio is
near unity, but there is a depression in the ratio along a track whose frequency
falls with increasing radius.  This dip seems to be 
due to an excess in fluctuation power in the accretion rate; in the case of
$r=3.5M$, a bump in $|\hat{\dot{M}}|^2$ can be clearly seen at
$\nu \simeq (0.2$--$0.4)\nuisco$ (Fig.~\ref{fig:lightcurves}).

The origin of
this excess can be identified by comparing its track in radius-frequency space
with the radial dependence of several significant dynamical frequencies: the
orbital frequency, the vertical epicyclic frequency, and the radial epicyclic
frequency.   As can be clearly seen in Figure~\ref{fig:emiss-mdot-psd-ratio},
the radial epicyclic frequency---but not the others---follows closely the centerline
of the dip, suggesting that radial epicyclic modes modulate the 
accretion rate without giving rise to emission---i.e. the oscillations are 
either not dissipated significantly, not dissipated locally, or both.  A similar
feature in the power spectrum of radial velocity as a function of radius was
noted by \cite{ReynMill09} in their data from a pseudo-Newtonian global disk
simulation.
The accretion rate also has more fluctuation power than the 
emissivity for $r>\risco$ and $\nu \simeq 0.3 \nuisco$.  We do not understand
the origin of this excess.  The net result of both excesses, however, is to
make $|\hat{\dot{M}}|^2$ a flatter function of frequency than ${|\hat {\cal L}}|^2$
at most radii and also to create deviations from power-law behavior in the
local accretion rate power spectra.

\begin{figure}
\plotone{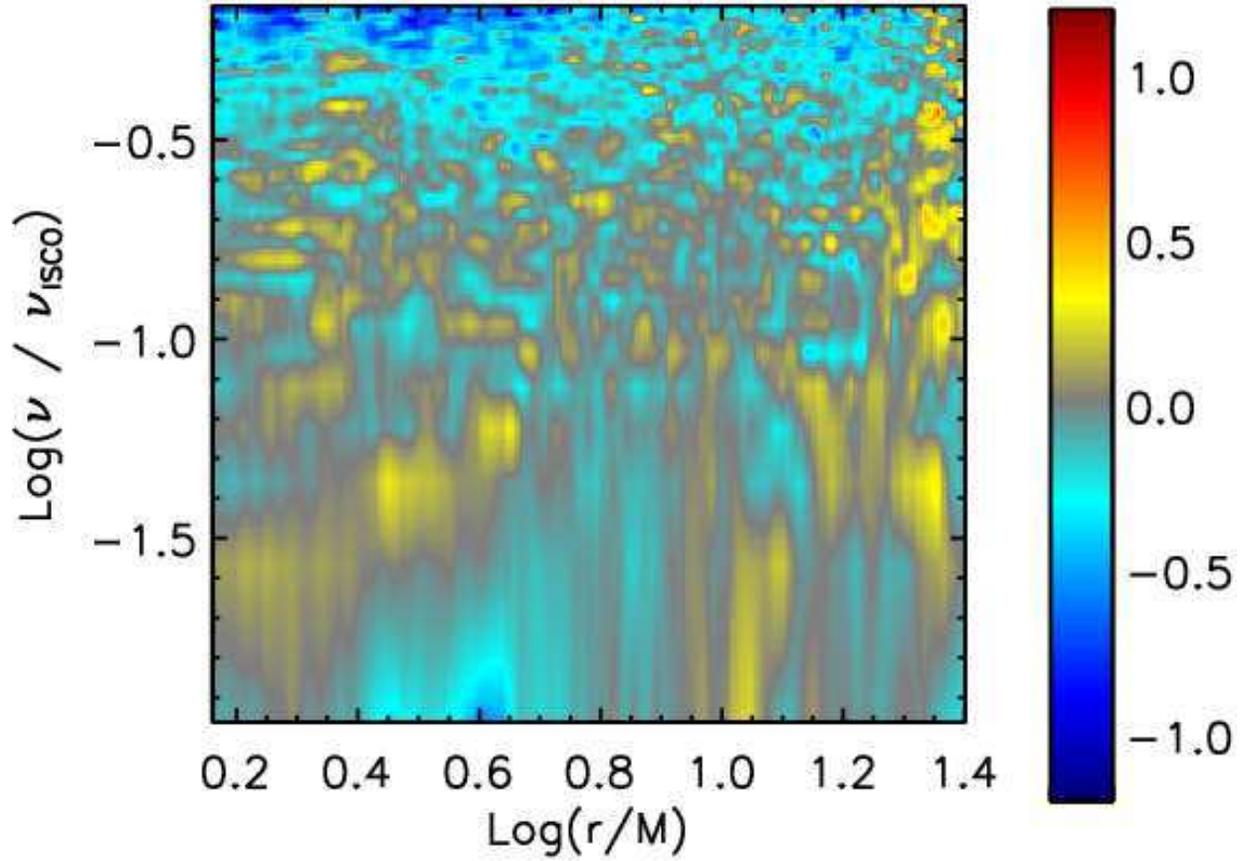}
\caption{Ratio of the normalized power spectrum of $dF/dr$ to the normalized power spectrum
of the emissivity as a function of radius and frequency.  Each power spectrum
was smoothed over nine frequency bins before the ratio was taken in order 
to display trends in the data more clearly. 
\label{fig:flux-emiss-psd-ratio}}
\end{figure}

We next study how closely fluctuations in ${\cal L}(r)$ are
mirrored in $dF/dr$.  To do so, we look at the ratio
\beq{
\mathcal{R} = \left( \frac{ \int \mathcal{L}(r,t) dt }{ \int \frac{dF(r,t)}{dr} dt } \right)^{2} 
\frac{\left| \widehat{\frac{dF}{dr}}(r,\nu) \right|^2}{ \left| \hat{\mathcal{L}}(r,\nu) \right|^{2}} 
\quad . \label{ratio-of-dFdr-emissivity-powers}
}
This is the ratio of the two normalized power spectra as a function of
radius and frequency
(Fig.~\ref{fig:flux-emiss-psd-ratio}).  We find that $\mathcal{R}$
is evenly distributed about unity, with deviations that rarely exceed a factor 
of $2$ in either direction.  Thus, 
the emissivity at $r$ predicts $dF(r)/dr$ at $\vartheta=0$ quite well.

We focus next on how the individual annular contributions to the
flux $dF/dr$ sum to the total flux $F(t)$.  One clue is given by the fact that
the fractional variances of $\mathcal{L}(r=3.5M)$ and $\dot{M}(r=3.5M)$
are rather similar, 0.152 and 0.175, respectively, while the fractional variance
of $F$ is rather smaller, 0.029.
We now understand that the emissivity follows the variability of the 
accretion rate (but with certain exceptions like those associated with radial
epicyclic motions) and $dF/dr$ varies like the emissivity.  
Why, though, does the total flux have such a small relative variance, and 
how can a set of oscillators (disk annuli) with power spectra 
that are $\alpha \sim -1.5$ power-laws integrate 
to have a composite PDS with $\alpha \simeq -2$?  

In the language of Section~\ref{sec:time-series-analysis}, the annuli can be
thought of as partitions with their own individual light curves.    
Since there are a large number of annuli or partitions, $S_b > S_a$ unless 
there is a dramatically low degree of phase-coherence between the different
radial segments.  If all the annuli were perfectly coherent,
$\Delta\psi_{n m} = 0 \, \forall n,m$, $P \simeq S_b$ and 
the light curve would have a $\alpha \sim -1.5$ power-law
power spectrum with a larger variance.  On the other hand, if all the annuli were 
completely incoherent, the total flux power spectrum would still be a power-law
with $\alpha \sim -1.5$, but with a smaller variance.   The only way to
steepen the slope of the spectrum is for the degree of coherence to decline with
increasing frequency.

\begin{figure}
\plotone{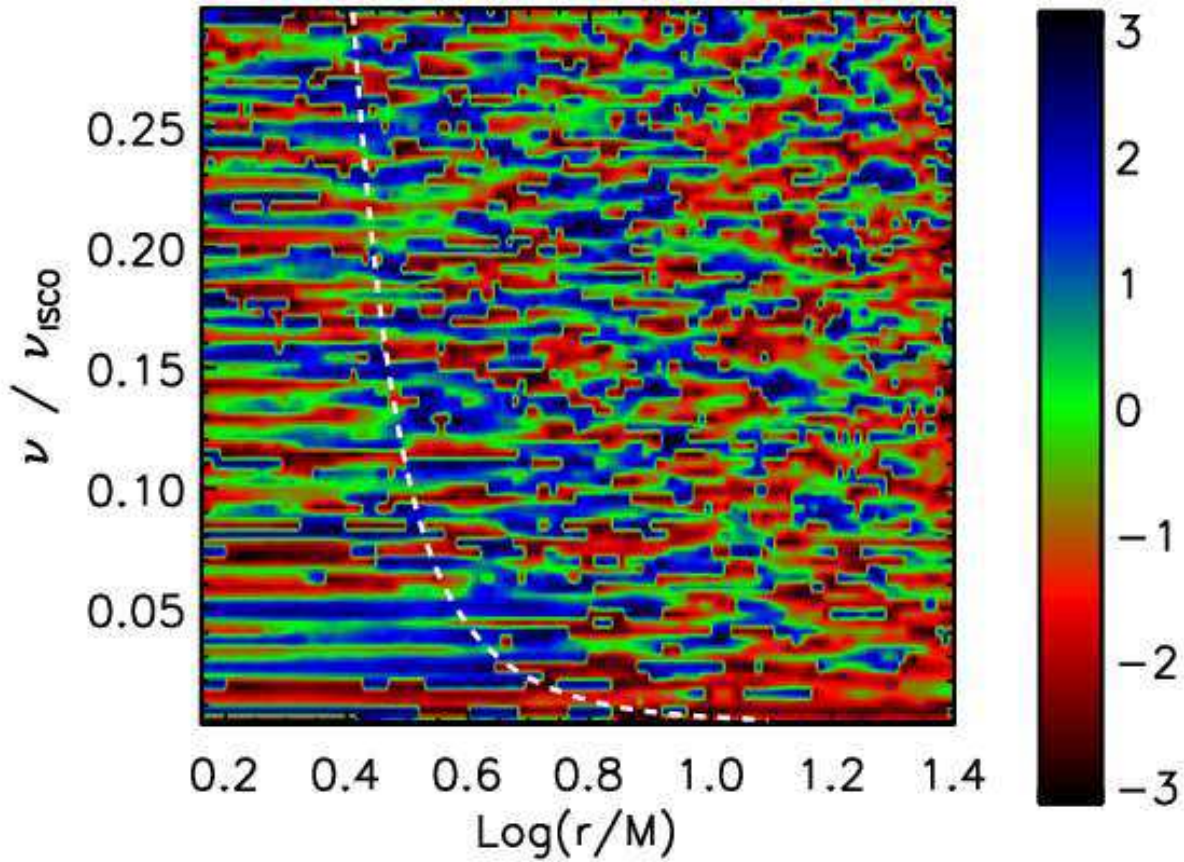}
\caption{(Left) Phase $\psi(\nu,r)$ of $dF/dr$ fluctuations when seen face-on 
($\vartheta=5^\circ$) for $\dot{m}=0.001$.  Note that we deviate from 
prior figure layouts and use a linear frequency scale here in order to resolve 
small-scale features.  In addition, we show only the lower half of our
frequency range.  The dashed curve is the local inflow rate
$\nu_\mathrm{inflow}$. (Right) The linear, periodic color map 
used to generate this figure. 
\label{fig:dfdr-phase}}
\end{figure}

The level of coherence in the variability of $dF/dr$ is illustrated in 
Figure~\ref{fig:dfdr-phase}, where we plot $\psi(\nu,r)$ over the lower half
of our frequency range.  At almost all radii,
$\psi(\nu,r)$ is incoherent in frequency (negligible correlation lengths in $\nu$),
but at fixed frequency, there can be significant coherence in radius.  The phases
are sufficiently coherent between different annuli that $S_b \gg S_a$, but their
correlations follow no simple pattern.  Different frequencies show
different radial coherence patterns, making it
impossible to state that radius $r$ varies coherently with radius $r^\prime$;
rather, one can only say that certain modes at $r$ are coherent with those at
$r^\prime$.

The white dashed curve in Figure~\ref{fig:dfdr-phase} shows the inflow rate
$\nu_\mathrm{inflow}$ as a function of radius, which we have defined to be the
mass-weighted mean radial velocity of bound material divided by the local
radial coordinate:
\beq{
\nu_\mathrm{inflow} r =  
\frac{\int_\mathrm{bound} \, d\theta \, d\phi \, dt \sqrt{-g} \rho u^r} 
{\int_\mathrm{bound} \, d\theta \, d\phi \, dt \sqrt{-g} \rho} 
\quad . \label{v-inflow}
}
A fluid element is considered bound if $h u_t > -1$, and 
$h$ is the fluid element's specific enthalpy. The time integral is performed 
over our standard epoch of $t=[7000M,15000M]$.  For $7M \lesssim r \lesssim 20M$,
we find that $\nu_\mathrm{inflow}(r) \simeq [28 T_\mathrm{orb}(r)]^{-1}$.  At
smaller radii, the inflow accelerates until near the ISCO and in the plunging
region $\nu_\mathrm{inflow}(r) \sim \Omega_K$.  Regions
to the left of this curve are clearly more coherent than those to the right.
That this should be so is not too surprising, given the ultimate dependence of
energy release on mass inflow. Indeed,
\citet{1997MNRAS.292..679L} proposed that the inner disk's low frequency
variability can be entirely explained by variations spawned at larger radii
(by fluctuations in the stress to pressure ratio)
that are then advected inward with the accretion flow.  What is demonstrated
in this phase picture is that fluctuations lower than the local inflow rate
do indeed propagate coherently inward, whatever their initial source.  However,
over much of the range of frequencies studied here, this criterion can be
satisfied only near the ISCO and in the plunging region itself.  At these
higher frequencies (which, as we shall see in the next section, are often
the object of most observational study), no such regular propagation pattern
can be discerned.  

Returning to the question of why the power spectrum of the total flux is steeper
than that of the flux radiated by individual annuli, we now see that this can
be explained by the diminution of the coherent radial range with increasing
frequency shown in Figure~\ref{fig:dfdr-phase}.


\subsection{Finite Speed of Light}
\label{sec:finite-speed-light}

\begin{figure}
\plottwo{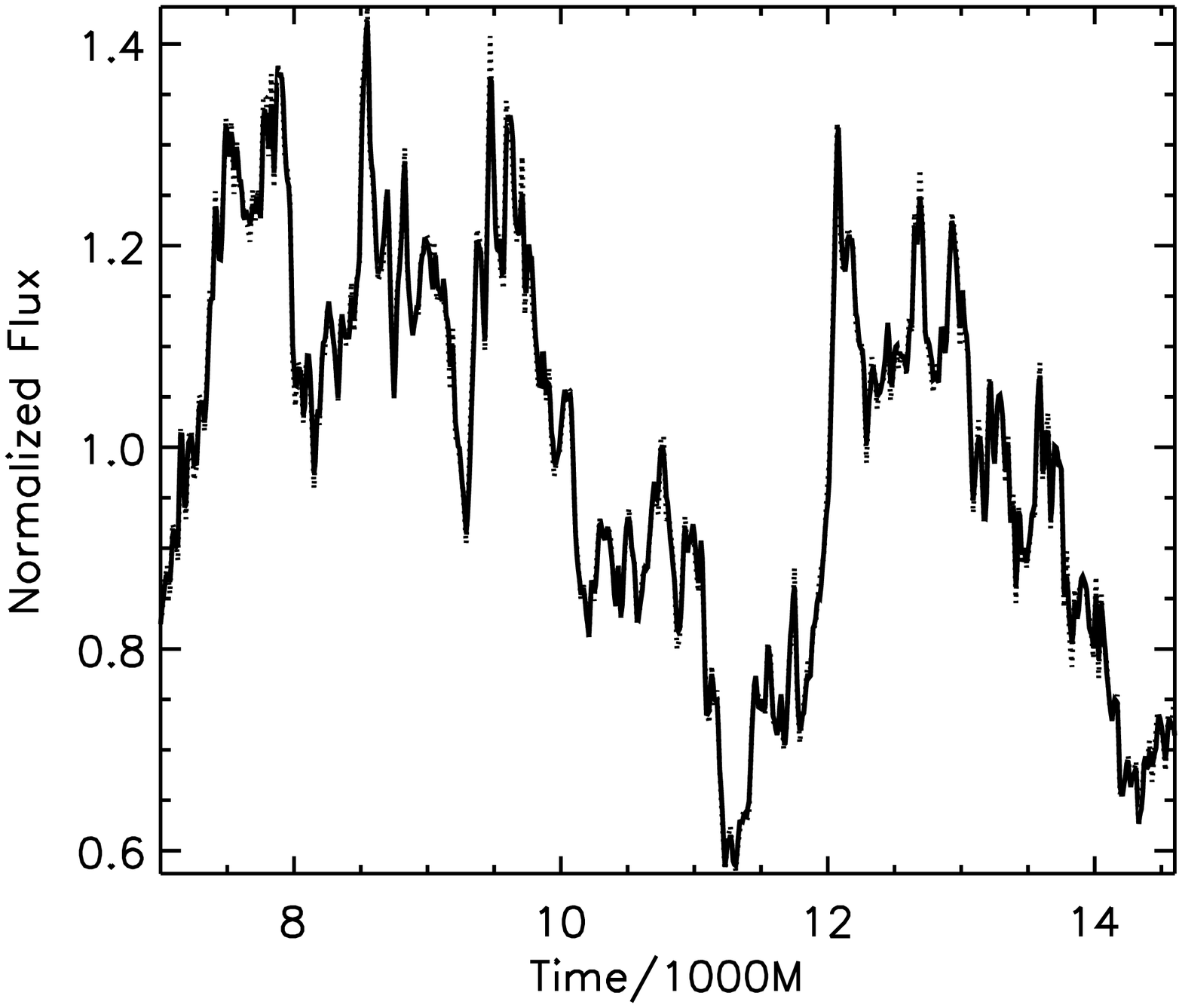}{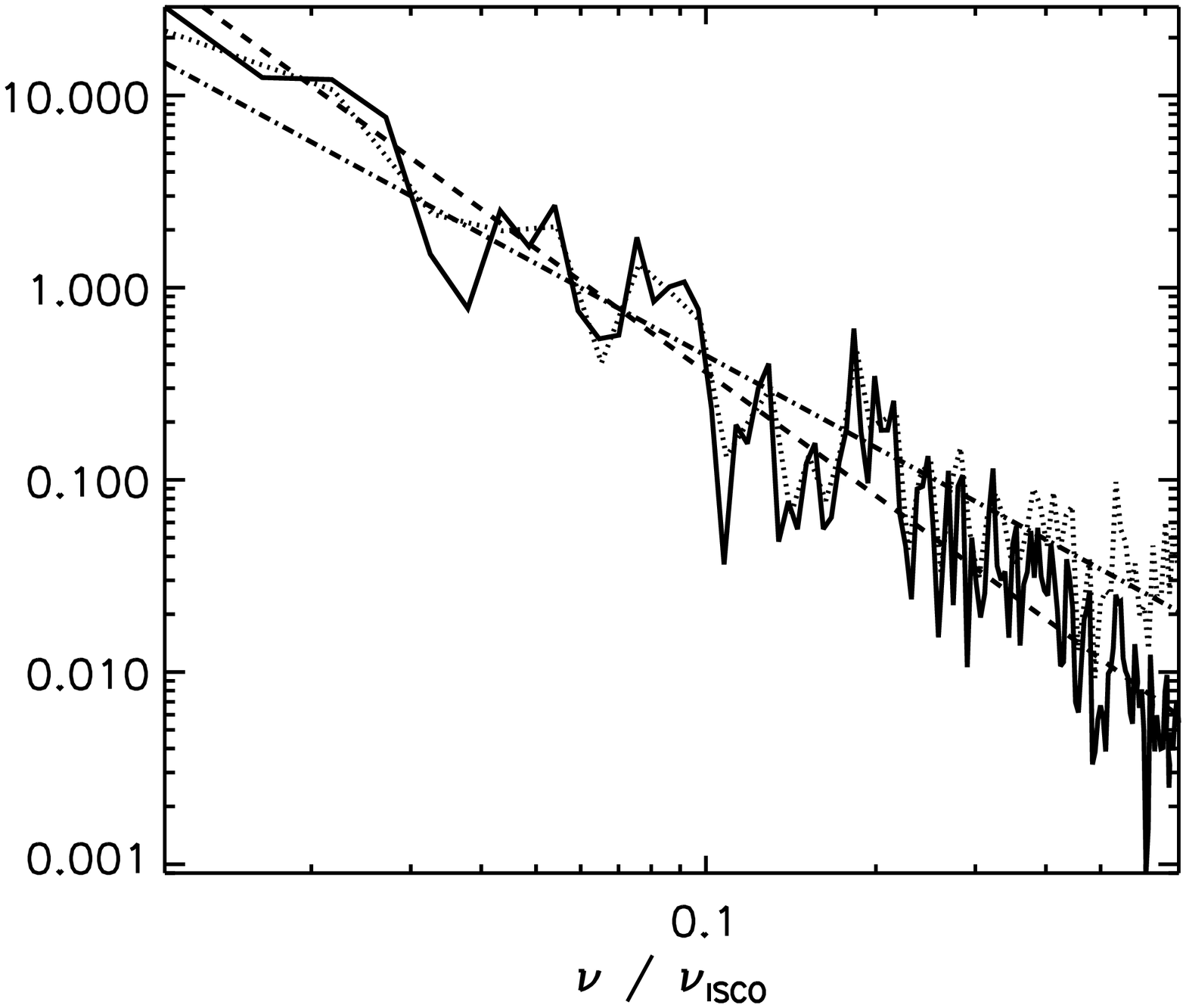}
\caption{(Left) Normalized light curves obtained from the calculation ignoring
time delay effects (dotted curve) and taking them into account (solid curve).
(Right) Normalized power spectra of these light curves compared to their 
best power-law fits; the dashed line represents the best fit to the data
with time delay effects, the dash-dot to the data in which time delays were
ignored.  Both light curves are for $\vartheta=29^\circ$ and $\dot{m} = 0.01$. 
\label{fig:light-curves-timedelay}}
\end{figure}

Our calculation for the first time correctly accounts for 
time delays while ray-tracing 3D GRMHD simulation data.  We would therefore
like to quantify how our results are affected by inclusion of this effect.
The light curves and power spectra from calculations with and without 
time delays are shown in Figure~\ref{fig:light-curves-timedelay}.   The light 
curves are identical except that the time delay calculation shows slightly
less short timescale variation.  This fact is illustrated more clearly in
the power spectra panel of this figure, which clearly shows that the fluctuation
power at high frequencies is diminished when one includes time delays.  

This contrast is easily understood.  Delay effects can diminish
coherence in the received signal when a region whose light crossing
time is $\Delta t$ varies coherently on timescales shorter than $\Delta t$.
On the other hand, enhancement of coherence by delay effects would require
remarkable contrivance because spatial and temporal fluctuations in the
turbulence would have to be correlated with the ray trajectories for
particular observers.   Consequently, photon time delays in general
decrease the fluctuation power.  The depression of the fluctuation power is
confined to the highest frequencies because maintenance of emissivity coherence
requires a coordinating signal propagating across the region, but
all signals, whether conveyed in bulk fluid motion or by some wave mode,
are limited to traveling no faster than $c$.  It follows that, for light
travel time effects to suppress variability, the coordinating signals must
be relativistic.  In the context of an accretion flow, relativistic signals
are largely confined to the innermost regions, which dominate the generation
of high frequency fluctuations.

\begin{figure}
\plotone{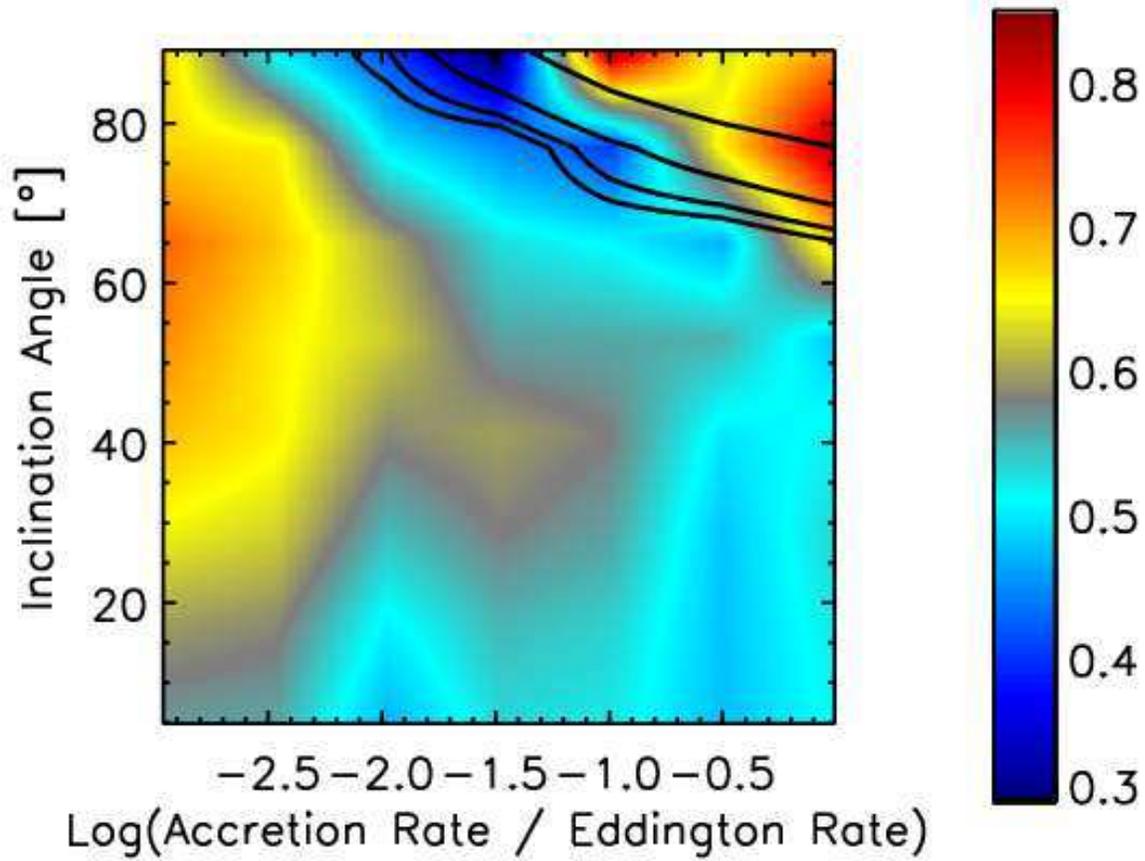}
\caption{Difference between the power-law exponents $\alpha$ from the calculation
without time delays to that with time delays as a function of 
$\dot{m}$ (horizontal axis) and $\vartheta$ (vertical axis).
The  cases in the upper-righthand corner of the 
plot are heavily obscured.  The black contour curves there 
represent---respectively---from bottom to 
top $R_o(\vartheta,\dot{m}) = \left\{ \risco, 3.5M, 6M, 12M\right\}$.
\label{fig:difference-power-law-exponents-pspace}}
\end{figure}

Because the time delay effect depends on the light's 
path through the material and the local velocity of the fluid, one expects 
it to depend on $\vartheta$ and $\dot{m}$.   We characterize its trend 
over parameter space in Figure~\ref{fig:difference-power-law-exponents-pspace},
where the difference in power-law exponents between the calculation with 
time delays and that without time delays is plotted.  In all cases, the time 
delay calculation yields a steeper PDS.  The contrast depends most strongly on
accretion rate, in the sense that it diminishes as the disk becomes more opaque;
this trend is consistent with the observation that as $\dot m$ increases, the
inner portion of the disk becomes progressively 
more obscured and contributes less to the power spectrum.  Larger $\vartheta$
produces slightly larger deviations between the two methods.
As the inclination angle increases, photon rays become more
nearly parallel to the disk's orbital velocity.  
For those fluid elements with relativistic velocities, the result is that 
fluid elements' worldlines move closer to the lightcone, leading
to somewhat greater coherence of emissivity along the rays.

\subsection{Azimuthal Symmetry Condition}
\label{sec:azim-symm-cond}

In order to save computational resources, we assumed in our simulation 
that the disk is periodic in $\phi$ on intervals equal to $\pi/2$.
Any modes longer than this are not included in our calculation, and the symmetry
condition artificially gives rise to correlations at this scale. The question then
arises:  are the absent modes and artificial symmetry important to 
our prediction?  

Even though we cannot rigorously evaluate the constraint's ramifications without 
repeating our analysis for a run on the entire domain, some insight might 
come from a similar simulation that used the full 
azimuthal extent \citep{2006ApJ...651.1031S}.  Their calculation 
used a numerical code that inadvertently preserves near-constant aspect 
ratio by failing to capture all dissipation as heat \citep{dVH03}.  
Even though their numerical techniques were different and no explicit cooling 
was used, our calculations share nearly identical initial conditions (besides the 
full azimuthal extent) and yield similar disk thicknesses.  Since the disk's
thickness dictates the poloidal size of turbulent eddies in the bulk, we may
expect that the characteristics of their correlations in $\phi$ will be applicable
to our system.  
They found that the surface density's dominant azimuthal correlation lengthscale is 
approximately $0.4\pi$, suggesting that our grid may be large enough to include
the most important modes.  

To quantify the systematic effect of the $\pi/2$ periodicity, we can employ
the partition formalism previously introduced.  For the purpose
of this discussion, it is convenient to label the quadrants by their panel
labels in Figure~\ref{fig:images-phi}, i.e., a, b, c, and d.
The quadrants are distinguished by the sign of their mean line-of-sight velocity
(receding or approaching) and their position (front or back).  Quadrants
a and c are approaching, whereas b and d are receding; quadrants a and b are
in back of the black hole, whereas c and d are in front.  Because
relativistic effects dominate obscuration effects in determining the
characters of their light curves, it is easiest to think of the system in
the optically thin limit.  If only special relativistic effects applied, the
two receding quadrants would produce identical light curves, as would the two
approaching quadrants.  However, general relativistic frame-dragging and light
deflection complicate the story.  For example, light in a particular direction
in the fluid frame can be wrapped around  the black hole and escape at a completely
different angle.  Although the deflection angle is large only very near the black
hole, most of the light is produced at these same radii, so it can be a very
important effect.  For instance, looking at Figure~\ref{fig:images-phi} 
we find that the most intense part of quadrant b is located on the 
opposite side of the black hole in the image plane because the brightest 
light---that which is emitted along the orbital velocity---has been bent 
around the black hole and focused toward the observer.  This phenomenon
means that quadrant b is more like an ``approaching'' quadrant
than a receding one.  
 
With these points in mind, we can now explain what controls the distinctions in
flux and power spectrum between the different quadrants.  For face-on views, they
all contribute identically to the light curve; as the viewing angle moves off-axis, 
special relativistic beaming and boosting enhances the approaching sides,
while general relativistic light bending and frame-dragging enhance the back sides.
The result is that over most of $\dot m$--$\vartheta$ parameter 
space, quadrant a is the brightest (both approaching and in back
of the black hole), d is the faintest (both receding and in front),
and b and c are similar to one another (b is receding but in back; c is approaching
but in front).  The maximum flux contrast between the brightest and
dimmest quadrants never exceeds a factor of $\sim 5$.  The slopes of their 
power spectra follow the same trend seen in flux: 
$\alpha_a \gtrsim \alpha_b \simeq \alpha_c > \alpha_d $ on average, with no 
spectral slope falling outside the range $-2.4 \le \alpha \le -1.8$.  
However, relative to quadrant c, the quadrant b becomes 
brighter and its PDS flatter as $\vartheta$ increases.
In the summed light curve, the contrasting effects largely cancel one another,
so that the spectral slope of the composite PDS can be described by a simple
average of the quadrants' individual power-law exponents.

In our calculation, the quadrants are precisely coherent at all frequencies
when viewed exactly face-on.  
As the inclination angle grows, they begin to
become incoherent at the highest frequencies, but even for $\vartheta = \pi/2$,
the range of incoherent frequencies is still quite limited.   The reason for
this behavior is that our symmetry condition makes their emissivity precisely
coherent, so such incoherence as exists is entirely due to time-delay
effects; as just discussed, they are small except at the highest frequencies.
Thus, if the absolute power spectrum from a single quadrant 
(before Doppler adjustments and obscuration effects)
is $A^2(\nu)$, our total power spectrum is $16A^2(\nu)$ when viewed on-axis,
and when viewed off-axis has essentially identical power at low frequencies, but
slightly less at high.

By contrast, in a full $2\pi$ simulation we expect that the emissivities of
the quadrants would have very similar power spectra to the emissivity we
calculate, but be completely incoherent if azimuthal correlations extend
only over angles $\simeq 0.4\pi$.  The same repertory of relativistic effects,
both special and general, will still apply, but we expect that they will similarly
cancel in sum.  Thus, a total flux power spectrum
$\simeq 4A^2(\nu)$ should result, as only the $S_a$ term contributes.  In other
words, if this reasoning holds, the shape of the power spectrum observed
from a full $2\pi$ disk would be quite similar to what we compute, but its
amplitude would be lower by about a factor of 4.

\section{Discussion and Conclusion}
\label{sec:discussion}

In this paper, we have presented a new, more physical method for 
estimating the temporal variability of radiation from the optically thin
(``coronal") regions of 3D GRMHD simulations. 
For the purpose of investigating
variability, it is necessary at this stage to separate optically thin regions
from optically thick because present-day global disk simulation codes do not
have the capacity to solve the transfer problem simultaneously with the
dynamics, and diffusion through optically thick regions materially alters
the character of variability.
The key improvement over previous calculations is the use of data from an
energy-conserving code with precise control of the disk's thermodynamics.
In addition, we have shown how to include photon travel
time delays, although they have a relatively small impact on the results
shown here.  We separated coronal emission from disk emission by integrating
the fluid emissivity from the scattering photosphere outward; the location of
the photosphere moves in a manner controlled by the nominal accretion rate in
Eddington units, $\dot m$.  Because our density data---which determines the 
optical depth---was written only every $20M$ in time over a period of $8000M$, 
we were constrained to explore the disk's power spectrum only over the frequency
range $\nu \in \left[3.5\times10^{-3}, 0.7\right]\nuisco$.

We found that the power spectrum of the observed flux's fluctuations in this
frequency band is described well by a
featureless power-law with index $\alpha \sim -2$ for essentially all optical
depths (or, in this formalism, accretion rates)
and inclination angles.  Although most of the fluctuation power has its
physical origin in accretion rate fluctuations, the slope of its power spectrum
is steeper by $\simeq 1$ than the slope of the accretion rate's power spectrum.
Two separate effects combine to create this steepening: there is high frequency
power in the accretion rate due to radial epicyclic motions that do not
contribute to variations in the emissivity; and the radial coherence of different
frequency modes declines with increasing frequency.  Thus, the power spectrum
of accretion rate fluctuations is {\it not} a good proxy for fluctuations in
the coronal light.  Because photon diffusion damping of high-frequency emissivity
fluctuations will likely steepen the power spectrum of the thermal luminosity,
we expect that the same will be true for the light curve of the thermal component.

Relativistic beaming and boosting cause the variance of the light curve to increase
with inclination angle, but do not materially change the shape of
the power spectrum.  The reason for this perhaps counter-intuitive result is that
Doppler effects flatten the power spectrum of the approaching segment of the
disk and steepen the power spectrum of the receding segment so that the two
changes compensate for one another.

Time-delay effects steepen the observed power spectrum at the highest frequencies.
If we had saved data from this simulation at intervals shorter than $20M$, we
expect that this effect would have been increasingly important at the higher
frequencies that would then have been accessible.  For this reason, in any
future work using simulation data to predict light curves on short timescales,
we strongly encourage proper accounting for photon travel times.

Our results span the frequency range $\sim 2.5\hz - 500\hz$ if scaled to
a $M=10\msun$ black hole and $\sim 2.5 \times 10^{-7}\hz - 5 \times 10^{-5}\hz$
for a $M=10^8\msun$ black hole.  Low frequency QPOs, which generally occur in
GBHs at $\sim 1$~Hz, would therefore be at best
only marginally detectable in our data.  High frequency QPOs, which are
sometimes seen at $\sim 200$--$300$~Hz, might in principle have appeared, but
we see no evidence for any.  On the other hand, they appear in real black
hole systems only in association with the transition to the steep power law
state.  Although
our simulation code very accurately conserves energy, the cooling function
we employ is no more than a toy-model.  A more complete description
of radiative cooling will be necessary to understand spectral state transitions,
and that might be a prerequisite for understanding high frequency QPOs as well.

Our optically thin limit ($\dot m = 0.001$) phenomenologically resembles the
hard state of Galactic black hole binaries in the sense that our definition of
``coronal" eliminates any optically-thick thermal disk in the inner part of the
accretion flow.  Intriguingly, the power-law
slope that we consistently find ($\alpha \simeq -2$) is crudely consistent
with the mean slope of the power spectrum measured in Cyg~X-1 in the range
1--500~Hz: steepening from $\simeq -1.7$ to $\simeq -2.4$ \cite{Rev00}.

For higher accretion rates, our corona is restricted to the outer layers
of the flow, more in keeping with what is often imagined for AGN.
A power-law slope $\simeq -2$ is also very roughly consistent with observations
of these objects.  For example, \cite{Mark09} shows that the slope of the
power spectrum in IC 4329A steepens from $\simeq -1$ to $\simeq -2$ across
the frequency range $10^{-8}$--$10^{-4}$~Hz.  Similarly, \cite{Mark07} find
that the power spectrum of Mrk~766 steepens from $\simeq -1.5$ to $\simeq -3$
from $\simeq 3 \times 10^{-5}$~Hz to $\simeq 10^{-3}$~Hz.  In this latter case,
Markowitz et~al. estimate that the central black hole mass may be only
$\sim 10^6$--$10^7 M_{\odot}$, which would place our simulated frequency range
roughly coincident with the observed band.

As already mentioned, the decrease in radial coherence length with
increasing frequency steepens the power spectrum of the aggregate light curve
relative to the power spectrum of the local emissivity.  In our very approximate
treatment, we described the result in terms of a new, steeper power-law.
A more careful and complete treatment might improve upon this description.
In particular, the factor that controls the radial coherence length is whether
the fluctuation frequency is larger or smaller than the local inflow rate.
It is the outward decrease of the inflow rate that leads to higher power
at lower frequencies by stretching the range of radial coherence.  However,
at sufficiently low frequencies, greater radial coherence does not add
appreciably to the power spectrum because material at larger radius does not
contribute much to the luminosity.  At frequencies lower than the
inflow rate at the radius within which most of the light is emitted, the
slope of the composite flux power spectrum should therefore match
the slope of the emissivity power spectrum.  One might then expect a smooth
roll-off from the slope of the emissivity power spectrum at these very
low frequencies to a steeper slope at high frequencies, similar to what is
observed in the hard states of GBHs and in AGN.  In a simulation, the
lowest reachable frequency
is the inverse of the duration; unfortunately, the radius within which our
simulation reached inflow equilibrium enclosed only a little more than half
the luminosity, so we did not reach low enough frequencies to see the change
in slope.

One should also be aware of several other caveats in evaluating these comparisons
with observations.
First, dissipation in magnetically-dominated plasmas is thought to 
entail particle acceleration across shocks or at reconnection sites.
These processes can be much more rapid than the orbital timescale, while the
inverse Compton cooling rate in units of $\Omega_K$ is
$\sim (m_p/m_e)(L_s/L_E)^{-1} (r/M)^{-1/2}$, for seed photon luminosity $L_s$.
Thus, when there is significant thermal disk radiation, the inverse Compton
cooling rate can likewise be much quicker than the orbital frequency.  Because
our cooling function has a characteristic rate $\sim \Omega_K$, it may
underestimate high frequency variability.   Second, our model focuses on the
total coronal luminosity, which is likely dominated by photons at energies
an order of magnitude higher than those generally studied in variability
observations.  If the power spectrum changes with photon energy (and there
are some hints of this: \cite{Mark07}), these may not be the appropriate
comparisons to make.  Third, our simulation did not distinguish the thermodynamic
properties of the disk body and the corona.  It is possible that a more
complete treatment of their thermal contrasts might alter the results.

In conclusion, we have presented a radiative cooling model, based directly
on simulations of 3D MHD turbulence
in general relativity, that predicts the power spectra of fluctuations in
hard X-ray flux observed from AGN and GBHs.  The calculation used 
a new ray-tracing procedure for correctly tracking the propagation of light through 
time and space within the time-dependent 3D GRMHD simulation data set.  The spectral 
slope found from our model---$\simeq -2$, significantly steeper than the slope
of the accretion rate power spectrum, depends only weakly on the inclination
and average accretion rate of the disk.  
Future simulations with more complete
physics and a more complete traversal of parameter space will shed further
light on this subject.

\acknowledgements
This work was supported by NSF grant AST-0507455 (JHK).  We thank 
Charles Gammie for his illuminating input regarding efficient 
bookkeeping algorithms for the time delay calculation. 
The computer resources of the 
Homewood High Performance Computer Cluster at Johns Hopkins 
University were used for our calculations.


\bibliography{bib}

\clearpage

\clearpage

\clearpage

\clearpage

\clearpage

\clearpage

\clearpage

\clearpage

\clearpage

\clearpage

\clearpage

\end{document}